\documentclass[byrevtex,pre,superscriptaddress,showpacs,showkeys,preprint,nofootinbib]{revtex4}
\usepackage{amsmath,amssymb}
\usepackage{epsfig}

\newcommand{\epl}{Europhys. Lett.}

\newcommand{\rcite}[1]{Ref.~\onlinecite{#1}}
\newcommand{\rcites}[1]{Refs.~\onlinecite{#1}}

\def\br{{\bf r}}
\def\bn{{\bf n}}
\def\ez{{\bf e}_z}
\def\er{{\bf e}_r}
\def\epsf{\varepsilon_F}
\def\epsp{\varepsilon_\Pi}
\def\epsm{\varepsilon_{\Pi_m}}
\def\hepsf{\hat{\varepsilon}_F}
\def\hepsp{\hat{\varepsilon}_\Pi}
\def\hsmen{\hat{S}_{\rm men}}
\def\vmen{V_{\rm men}}
\def\avsup{\mbox{}^aV_{\rm sup}}
\def\avcorr{\mbox{}^aV_{\rm corr}}
\def\bvsup{\mbox{}^bV_{\rm sup}}
\def\bvcorr{\mbox{}^bV_{\rm corr}}
\newcommand{\Ezp}{\hat{E}_{z,+}}

\newcommand{\Ep}{\hat{\bf E}_{\parallel}}

\begin{document}

\title{Theory of capillary--induced interactions beyond the
  superposition approximation}

\author{Alvaro Dom\'\i nguez}
\affiliation{F\'\i sica Te\'orica, Universidad de Sevilla, Apdo.\ 1065, E--41080 Sevilla, Spain}
\author{Martin Oettel}
\affiliation{Institut f\"ur Physik, Universit{\"a}t Mainz, WA 331, D-55099 Mainz, Germany}
\author{S.\ Dietrich}
\affiliation{Max--Planck--Institut f\"ur Metallforschung, Heisenbergstr.\ 3, D--70569 Stuttgart, Germany}
\affiliation{Institut f\"ur Theoretische und Angewandte
  Physik, Universit\"at Stuttgart, Pfaffenwaldring 57, D--70569 Stuttgart, Germany}

\date{June 1, 2007}

\begin{abstract}
  Within a general theoretical framework we study the effective,
  deformation--induced interaction between two colloidal particles
  trapped at a fluid interface in the regime of small 
  deformations. In many studies, this interaction has been computed
  with the ansatz that the actual interface configuration for the pair
  is given by the linear superposition of the interface deformations
  around the single particles. 
  Here we assess the validity of this approach and compute the
  leading term of the effective interaction for large interparticle separation 
  beyond this so-called superposition
  approximation. As an application, we consider the experimentally
  relevant case of interface deformations owing to the electrostatic
  field emanating from charged colloidal particles. In mechanical
  isolation, i.e., if the net force acting on the total system
  consisting of the particles plus the interface vanishes, the
  superposition approximation is actually invalid. The effective
  capillary interaction is governed by contributions beyond this
  approximation and turns out to be attractive. For sufficiently small
  surface charges on the colloids, such that linearization is strictly
  valid, and at asymptotically large separations, the effective
  interaction does not overcome the direct electrostatic repulsion
  between the colloidal particles.
\end{abstract}


\maketitle 

\section{Introduction}
\label{sec:intro}

The self--assembly of sub-$\mu$m colloidal particles at fluid (e.g.,
water/air or water/oil) interfaces has gained significant interest in
view of various basic and applied issues such as the study of
two--dimensional melting \cite{Pier80}, investigations of mesoscale
structure formation \cite{Joan01}, and engineering of colloidal
crystals on spherical surfaces \cite{DHNM02}.
The colloidal particles are trapped at the interface if the fluid
phases wet the colloid only partially; this configuration is stable
against thermal fluctuations and it appears to be even the global
equilibrium state, in accordance with the experimental observation that the
colloids immersed in the bulk phases are attracted towards the
interface \cite{Pier80}. 

In order to prevent coagulation, the colloidal particles are
electrically charged. The ensuing repulsive force is well understood
and at large separations $d$ it varies like a dipole--dipole
interaction $\propto 1/d^4$, because the monopoles vanish due to
screening by counterions in water \cite{Hurd85,ACNP00}.
Nevertheless, several experimental findings have led to postulating an
attractive effective force between such particles with a range much
larger than that of van--der--Waals forces
\cite{RGI97,GhEa97,SDJ00,QMMH01,GEFM01,MGIR02,NBHD02,TAKK03,GoRu05,CTHN06} (but see also \rcite{FMMH04}).
Spherical particles (radii $R=0.25\dots 2.5$ $\mu$m) at flat
water--air interfaces exhibit the spontaneous formation of complicated
metastable mesostructures consistent with the presence of a minimum in
the effective intercolloidal potential at separations $d/R\approx
3\dots 20$ and with a depth of at least a few $k_B T$.

\begin{figure}
  \begin{center}
    \epsfig{file=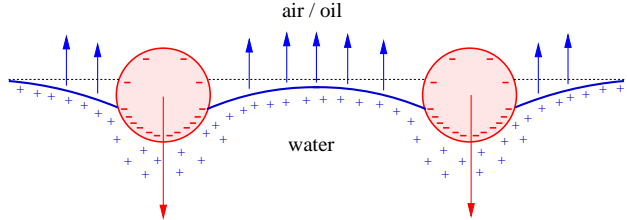,width=.5\textwidth}
    \caption{ Schematic drawing of the deformation of a fluid
      interface by electric fields due to charged colloidal particles
      trapped at the interface. Counterions gather on the side of the
      electrolytic phase (water) and a pressure field arises which
      pulls on the interface and on the particles.}
    \label{fig:2coll_side}
  \end{center}
\end{figure}

Until now, no unequivocal explanation for the appearance of these
relatively long--ranged attractions is available.  One possibility,
which has been explored intensively in previous years, consists of an
attraction mediated by the deformation of the interface (see
Fig.~\ref{fig:2coll_side}). This is similar to the so-called flotation
force attracting particles floating at the surface of water which is
deformed by the weight of the particles \cite{Nico49,CHW81}. But
gravity plays no role for micrometer sized particles as described
above. Instead, the electrostatic field around the charged colloids
deforms the interface and gives rise to effective, capillary--induced
interactions.
There have been contradictory results about the properties of this
effective interaction. In \rcites{NBHD02,DKB04} it has been 
argued in favor
of an attractive force decaying like $1/d$ (similar to the
gravity--induced flotation force). In
\rcites{MeAi03,FoWu04,ODD05a,DOD05,ODD06} this line of thought 
was shown to be invalid because the decay 
turns out to be
much faster if the physical system consisting of the colloidal
particles plus the interface is mechanically isolated, i.e., if the
total force acting on this system vanishes in the limit of a
macroscopically extended interface with negligible borders. This is in
principle the case for the experiments conducted in Langmuir troughs
with lateral extensions several times the capillary length of the
interface (i.e., orders of magnitude larger than the colloid radius).
In this case the electrostatic force pulling on the interface is
counterbalanced exactly by the electrostatic force pushing the
particles into water.
In particular, the authors of \rcite{MeAi03} argue for an attractive
force decaying like $1/d^7$. However, this was in turn corrected in
\rcites{FoWu04,ODD05a} and a repulsive force with an asymptotic
decay $\propto 1/d^7$ was derived. In \rcite{ODD05a} it was noticed
that this latter result is actually unreliable because the linear 
superposition approximation employed in these calculations is not valid
if mechanical isolation holds. Furthermore, since it is the electric
field ${\bf E}$, and not the electrostatic pressure $\propto {\bf
  E}^2$, which obeys a superposition principle, the superposition
approximation will be unsuitable if the main contribution to the
electrostatic pressure stems from cross terms in ${\bf E}^2$. This
latter case was studied in \rcites{ODD05b,WuFo05} with the
conclusion that the capillary--induced force is attractive and decays
like $1/d^4$, which is the same asymptotic behavior as the direct
dipole--dipole repulsion; whether the capillary attraction overcomes
the electric repulsion must then be determined by a detailed analysis
of the electrostatic problem. The total force is repulsive in the
regime of small deformations of a flat interface (equivalent to the
regime of small colloidal charges), for which the calculations were
carried out \cite{ODD05b,WuFo05}. Only for sufficiently large
colloidal charges the capillary attraction may asymptotically overcome
the electrostatic repulsion leading to a minimum in the total
effective potential \cite{ODD05b}. However, further calculations ---
going beyond the linearization assumptions in treating the
electrostatic pressure on the interface and the energy of the deformed
interface --- are necessary to substantiate this claim.
The experiment described in \rcite{NBHD02} is peculiar in the sense that
the unperturbed interface is actually that of a relatively small
spherical droplet pending from a plate. The importance of this
finite--size effect was studied in
\rcites{ODD05a,DOD05,DOD06b,Wuer06b,DOD07a}.  In
\rcites{DOD05,DOD06b,DOD07a} it was found that the flotation--like
decay $1/d$ is present because the plate breaks the condition of
mechanical isolation, but it is quantitatively too small to explain
the experimental observations\footnote{In this respect, \rcite{ODD05a}
  is incomplete but the same conclusion about the
  irrelevance of this effect is reached. \rcite{Wuer06b} contains
  several important mathematical errors \cite{DOD07a} 
  and the corresponding conclusion erroneously disagrees with the 
  actual irrelevance of the finite--size effect.}.

Here we follow the approach of \rcite{ODD05a} in order to calculate
the capillary--induced effective interaction beyond the superposition
approximation in the case that there is a pressure field of general form
$\hat{\Pi}(\br)$ acting on the interface in the limit of small
deformations. The limit of small deformations corresponds to an
analysis to leading order in the small dimensionless parameters
$\hepsf$ and $\hepsp$, defined in Eq.~(\ref{eq:epsfp}) below, 
which are measures of the force
acting on the colloidal particles and the interface, respectively. 
In this manner, inter alia 
we provide a mathematically sound derivation of the results reported
in \rcites{ODD05b,WuFo05}. In Sec.~\ref{sec:theory} we derive the
exact leading--order expressions for the capillary potential and show
in which respect the superposition approximation becomes inconsistent
if mechanical isolation holds. In such a case, the final result
depends 
on the stage at which the approximation is introduced and applied. In
Subsec.~\ref{sec:asymptotic} we derive the leading asymptotic dependence
on $d$ of the effective interaction using only some rather general
assumptions on the form of $\hat{\Pi}(\br)$. The result for the
effective potential energy is summarized in Eq.~(\ref{eq:vmen_asymp}).
In Sec.~\ref{sec:appl} we analyze the particular case that the
pressure field $\hat{\Pi}(\br)$ is due to the electric field created
by charged colloidal particles in a mechanically isolated system. We
consider two limiting cases in order to solve the electrostatic
problem: (i) water as a perfect conductor (i.e., vanishing Debye
length), and (ii) the colloidal particles as point--like objects (i.e.,
vanishing particle radius). In both cases we recover the conclusion
that the effective interaction is attractive but can overcome the
electric repulsion only if $\hepsf\gtrsim 1$, which is outside the
small--deformation regime considered here.
Finally, in Sec.~\ref{sec:end} we summarize our results and discuss
their relevance in connection with the experiments described in the
literature.

\section{Free energy of effective capillary interaction}
\label{sec:theory}

\subsection{Exact results}
\label{sec:exact}

\begin{figure}
  \begin{center}
    \epsfig{file=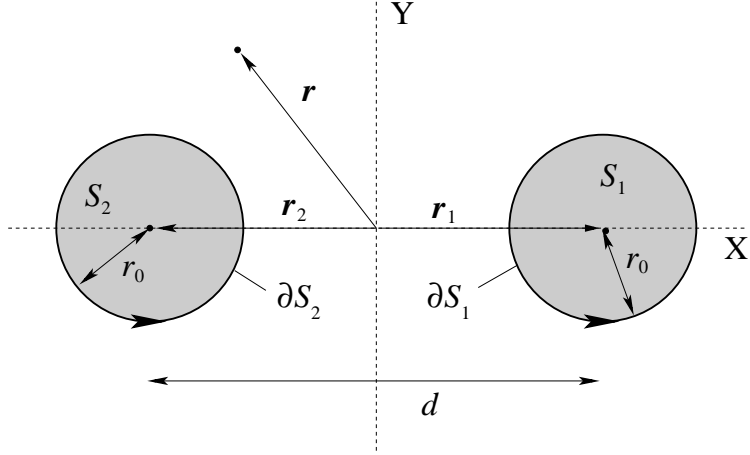,width=.6\textwidth}
    \caption{ Top view (plane $z=0$) of the reference configuration
      with two colloids. $d$ is the fixed lateral distance between 
      the colloid centers projected onto the plane $z=0$. 
      $S_1$ and $S_2$ are disks
      of radius $r_{0}$, the corresponding circumferences
      (counterclockwise) are $\partial S_1$ and $\partial S_2$. The
      (projected) interface is 
      $\hsmen = \mathbb{R}^2\backslash(S_1 \bigcup S_2)$.
      The position of any point on the plane is denoted with $\br$; in
      particular, $\br_\alpha$ is the position of the 
      colloid $\alpha$.
    }
    \label{fig:2coll_top}
  \end{center}
\end{figure}

We consider $N$ identical\footnote{In actual experiments there is a
  certain degree of polydispersity in size, shape, electric charge,
  etc., which we do not expect to alter the conclusions qualitatively
  if it remains small enough.}  spherical particles trapped at a fluid
interface (see Fig.~\ref{fig:2coll_top} for $N=2$). We define the
reference configuration as a flat interface in the plane $z=0$ and the
colloids at a height such that the colloid--interface contact occurs
at Young's contact angle $\theta \in (0,\pi)$. In this configuration,
$S_\alpha$ denotes the circular disk delimited by the contact line on
the colloid $\alpha$, $\partial S_\alpha$ is the corresponding contact
line (of radius $r_0 = R \sin\theta$ for a particle of radius $R$) 
traced counterclockwise when viewed from the
top\footnote{Which fluid phase is to be the top one depends on the
  experimental setup. Usually gravity breaks the up--down
  symmetry.}, and $\hsmen$
is the fluid interface (with surface tension $\gamma$), enclosed by a
boundary $C_L$ of typical size of the order of $L$ representing, e.g.,
the vessel containing the system.
The relative lateral positions of the colloids are kept fixed and thus we
consider only vertical displacements $\hat{u}(\br=(x,y))$ of the fluid
interface and of the height $\Delta \hat{h}_\alpha$ of the center of
the colloid $\alpha$ relative to the plane $z=0$. In the reference
configuration there is a (vertical) force $\hat{F}_\alpha$ acting on
the colloid $\alpha$ and a (vertical) force per unit area
$\hat{\Pi}(\br)$ on the meniscus. We define the dimensionless forces
\begin{equation}
  \label{eq:epsfp}
  \hat{\varepsilon}_{{F}_\alpha} := - \frac{\hat{F}_\alpha}{2 \pi \gamma r_0} , \qquad 
  \hepsp := \frac{1}{2 \pi \gamma r_0} 
  \int_{\hsmen} \!\!\!\!\! dA \; \hat{\Pi} ,
\end{equation}
where $dA$ is the element of interface area.
The reference configuration is the equilibrium state in the absence of
forces, $\hat{F}_\alpha \equiv 0, \hat{\Pi} \equiv 0$.  Within the
approximation of small deviations from the reference configuration,
the free energy of the system with respect to this configuration is
\cite{ODD05a}
\begin{equation}
  \label{eq:freeF}
  {\cal \hat{F}} =
  \gamma \int_{\hsmen} \!\!\!\!\!\! dA \; 
  \left[ \frac{1}{2} |\nabla \hat{u}|^2 
    - \frac{1}{\gamma} \hat{\Pi} \, \hat{u} \right] + \sum_{\alpha=1}^N \left\{
    \frac{\gamma}{2 r_{0}} \oint_{\partial S_\alpha} \!\!\! d\ell \; [\Delta \hat{h}_\alpha - \hat{u}]^2 
    - \hat{F}_\alpha \Delta \hat{h}_\alpha \right\} 
  + {\cal O}(\hepsf,\hepsp)^3,
\end{equation}
where $d\ell$ is the element of arclength.
The free energy contains a contribution from the change of the contact
area between the phases (two fluid phases and the solid particles),
and a contribution from the work done by the forces $\hat{F}_\alpha$
and $\hat{\Pi}$ via displacements from the reference configuration.

The values of $\hat{u}({\bf r})$ and $\Delta \hat{h}_\alpha$ in the
equilibrium state are determined by minimizing this free energy.  This
leads to the following equations:
\begin{subequations}
  \label{eq:equil}
\begin{align}
  \label{eq:equil_h}
  \Delta \hat{h}_\alpha &=
  \langle{\hat{u}}\rangle_\alpha - 
  r_0 \, \hat{\varepsilon}_{{F}_\alpha} , 
  & &\langle{\cdot}\rangle_\alpha := \frac{1}{2 \pi r_{0}} 
  \oint_{\partial S_\alpha} \!\!\! d\ell \; (\cdot) , \\
  \label{eq:equil_YL}
  \nabla^2 \hat{u} &= 
  - \frac{1}{\gamma} \hat{\Pi} , 
  & &\br \in \hsmen \\
  \label{eq:equil_bc}
  \bn_\alpha \cdot \nabla \hat{u} ({\bf r}) &=
  \hat{\varepsilon}_{{F}_\alpha} + 
  \frac{\hat{u}({\bf r})-\langle{\hat{u}}\rangle_\alpha}{r_{0}} , 
  & &\br \in \partial S_\alpha , \\
  \label{eq:equil_pin}
  \hat{u}({\bf r}) &= 0 , & &\br \in C_L ,
\end{align}
\end{subequations}
where $\bn_\alpha$ is the unit vector in the outward
normal direction of $\partial S_\alpha$. 
Equation~(\ref{eq:equil_h}) is a geometrical relationship,
Eq.~(\ref{eq:equil_YL}) describes local mechanical equilibrium (the
pressure $\hat{\Pi}$ is compensated by the curvature--induced
interfacial tension), Eq.~(\ref{eq:equil_bc}) describes mechanical
equilibrium of the particle (the force $\hat{F}_\alpha$ is balanced by
the interfacial tension exerted at the contact line), and 
Eq.~(\ref{eq:equil_pin}) represents a boundary condition at the
external border $C_L$ (for simplicity we take a pinned interface, but
of course other physically reasonable boundary conditions are
possible, the details of which are actually irrelevant in the limit
$L\to\infty$ we shall consider \cite{ODD05a}).
The free energy functional in Eq.~(\ref{eq:freeF}) {\em evaluated at
  the equilibrium configuration} can be simplified by using the
relationships in Eq.~(\ref{eq:equil}) and we obtain two equivalent
expressions:
\begin{subequations}
\label{eq:equilF}
\begin{eqnarray}
  \label{eq:freeF1}
  {\cal \hat{F}}_{\rm eq} & = & 
  - \frac{1}{2}\gamma \int_{\hsmen} \!\!\!\!\!\! dA \; 
  |\nabla \hat{u}|^2 
  + \pi \gamma \sum_{\alpha=1}^N \left[
    \langle\hat{u}\rangle_\alpha^2 - \langle\hat{u}^2\rangle_\alpha
    - r_0^2 \, \hat{\varepsilon}_{{F}_\alpha}^2 
  \right]  \\
  \label{eq:freeF2}
  & = & - \frac{1}{2} \int_{\hsmen} \!\!\!\!\!\! dA \; 
  \hat{\Pi} \, \hat{u}
  + \pi \gamma \sum_{\alpha=1}^N r_0 \, \hat{\varepsilon}_{{F}_\alpha} 
   \left[ \langle\hat{u}\rangle_\alpha 
     - r_0 \, \hat{\varepsilon}_{{F}_\alpha} 
  \right] .
\end{eqnarray}
\end{subequations}

From this point onward, we shall consider the particular case of two
identical colloids. We assume that any external force acting on the
system (e.g., gravity) is independent of the positions of the 
particles\footnote{This seems to be actually a reliable approximation for the
  experimental setups considered so far in the literature, but see the
  discussion in Sec.\ \ref{sec:end}.}, so that symmetry
arguments 
will allow one to simplify the expressions. 
We use the notations $u_\alpha ({\bf r}) := u(|{\bf r} - {\bf
  r}_\alpha|)$, $\Pi_\alpha ({\bf r}) := \Pi(|{\bf r} - {\bf
  r}_\alpha|)$, $\epsf := \hat{\varepsilon}_{F_\alpha}$, $S_{\rm men,
  \alpha} := \mathbb{R}^2\backslash S_\alpha$ for the corresponding
quantities in the presence of a {\em single} colloid located at
position ${\bf r}_\alpha$. This means that the function $u_\alpha({\bf
  r}) \; (\alpha=1, 2)$ satisfies
\begin{subequations}
  \label{eq:single}
\begin{align}
  \label{eq:single_a}
  \nabla^2 {u}_\alpha &=
  - \frac{1}{\gamma} {\Pi}_\alpha ,
  & &\br \in {S}_{\rm men, \alpha} , \\
  \bn_\alpha \cdot \nabla u_\alpha ({\bf r}) &=
  \epsf, 
  & &\br \in \partial S_\alpha , \\
  \label{eq:single_pin}
  {u}_\alpha({\bf r}) &= 0 
  & &\br \in C_L .
\end{align}
\end{subequations}
In terms of these single--colloid solutions the configuration
in the presence of two colloids can be written without loss of
generality as
\begin{subequations}
  \begin{align}
    \label{eq:hpi}
    \hat{\Pi} &= \Pi_1 + \Pi_2 + 2 \Pi_m , \\
    \label{eq:hu}
    \hat{u} &= u_1 + u_2 + u_m , \\
    \label{eq:hepsf}
    \hepsf &= \epsf + \varepsilon_m ,
  \end{align}
\end{subequations}
where we have introduced $\hepsf := \hat{\varepsilon}_{F_1} =
\hat{\varepsilon}_{F_2}$ reflecting the symmetry of the
problem.
The fields $u_m(\br)$ and $\Pi_m(\br)$ and the quantity
$\varepsilon_m$ introduced this way represent
the corrections to the so-called superposition approximation, which is
defined by setting
\begin{subequations}
  \label{eq:superpositionAnsatz}
\begin{eqnarray}
  {u}_m & = & 0 , \\
  {\Pi}_m & = & 0 , \\
  \varepsilon_m & = & 0 , 
\end{eqnarray}
\end{subequations}
i.e., the effects of other particles on the single--particle
configuration are neglected altogether.
From 
Eqs.~(\ref{eq:equil}, \ref{eq:single}) one can derive the following
equations linking
$u_m(\br)$, $\Pi_m(\br)$, and $\varepsilon_m$ 
(with $\alpha,\beta=1,2$):
\begin{subequations}
  \label{eq:cross}
\begin{align}
  \label{eq:cross_YL}
  \nabla^2 {u}_m &= - \frac{2}{\gamma} {\Pi}_m ,
  & &\br \in \hsmen , \\
  \label{eq:cross_bc}
  \bn_\alpha \cdot \nabla u_m 
  - \frac{u_m - \langle u_m \rangle_\alpha}{r_0} &=
  \varepsilon_m 
  - \bn_\alpha \cdot \nabla u_\beta
  + \frac{u_\beta - \langle u_\beta \rangle_\alpha}{r_0} ,
  & &\br \in \partial S_\alpha 
  \quad (\beta \neq \alpha) , \\
  {u}_m({\bf r}) &= 0 ,
  & &\br \in C_L .
\end{align}
\end{subequations}
The superposition approximation is violated even if $\Pi_m=0$ and
$\varepsilon_m=0$ because of the boundary conditions at the contact
lines (Eq.~(\ref{eq:cross_bc})), as pointed out in \rcite{ODD05a}. The 
case of non-vanishing $\Pi_m$ was addressed in
\rcites{ODD05b,WuFo05}.

The capillary--induced effective interaction energy is defined as $\vmen (d) :=
\hat{\cal F}_{\rm eq} - 2 {\cal F}_{\rm eq}$, where $2 {\cal F}_{\rm
  eq}$ is the sum of the equilibrium free energies of the 
single--colloid
configurations, i.e., for $d\to\infty$.
$\vmen(d)$ depends parametrically on the (lateral) separation $d$
of the colloid centers in the reference configuration. This is {\em
  not} the total interaction potential, which must include, e.g., the
direct electrostatic repulsion between charged colloids, not
considered in the expression~(\ref{eq:freeF}) for the free energy.
The effective interaction energy can be written as $\vmen = V_{\rm
  sup} + V_{\rm corr}$, where $V_{\rm sup}$ is the result of imposing the
superposition approximation (Eq.~(\ref{eq:superpositionAnsatz})),
and $V_{\rm corr}$ is the
correction to this approximation. If we use the
expression~(\ref{eq:freeF1}), we obtain
\begin{subequations}
  \label{eq:Vmen1}
  \begin{equation}
    \label{eq:Vsup1}
    \avsup =
    - \gamma \int_{\hsmen} \!\!\!\!\!\! dA \; 
    (\nabla u_1) \cdot (\nabla u_2)
    + \gamma \int_{S_1} \!\!\! dA \; |\nabla u_2|^2 
    - 2 \pi \gamma \left\langle u_2^2 - \langle u_2 \rangle_1^2 
    \right\rangle_1 ,
  \end{equation}
  \begin{align}
    \label{eq:Vcorr1}
    \avcorr =&
    - \frac{1}{2} \gamma \int_{\hsmen} \!\!\!\!\!\! dA \; 
    (\nabla u_m) \cdot \nabla (u_m + 4 u_2) 
    \nonumber \\
    & - 2 \pi \gamma \left\langle u_m (u_m + 2 u_2)
      - \langle u_m \rangle_1 \langle u_m + 2 u_2 \rangle_1 
    \right\rangle_1
    - 2 \pi \gamma r_0^2 \varepsilon_m ( 2\epsf + \varepsilon_m) .
  \end{align}
\end{subequations}
On the other hand, using expression~(\ref{eq:freeF2}), we
obtain
\begin{subequations}
  \label{eq:Vmen2}
  \begin{equation}
    \label{eq:Vsup2}
    \bvsup =
    - \int_{\hsmen} \!\!\!\!\!\! dA \; \Pi_1 u_2
    + \int_{S_1} \!\!\! dA \; \Pi_2 u_2
    + 2 \pi \gamma r_0 \, \epsf \langle u_2 \rangle_1 ,
  \end{equation}
  \begin{align}
    \label{eq:Vcorr2}
    \bvcorr =&
    - \int_{\hsmen} \!\!\!\!\!\! dA \; 
    \left[\Pi_2 \, u_m + 2 \Pi_m \, u_2 
      + \Pi_m \, u_m \right] 
    \nonumber \\
    & + 2 \pi \gamma r_0 \langle (\epsf + \varepsilon_m) u_m 
    + \varepsilon_m (u_1 + u_2) \rangle_1
    - 2 \pi \gamma r_0^2 \varepsilon_m ( 2\epsf + \varepsilon_m) .
  \end{align}
\end{subequations}
We emphasize that the two alternative expressions of $V_{\rm sup}$ or
$V_{\rm corr}$ are not equivalent (but their sum $\vmen$ is), and they
in turn differ from $V_{\rm sup}$ as computed in \rcite{ODD05a} (see
Eq.~(39) therein), which was derived by inserting the superposition
ansatz directly into Eq.~(\ref{eq:freeF}) here. Application of Gauss'
theorem with Eq.~(\ref{eq:single_a}) leads to
\begin{equation}
  \label{eq:discrepancy}
  \avsup - \bvsup = 2\pi\gamma \left\langle r_0 (u_1 + u_2) 
    \frac{\partial u_2}{\partial n_1} - 
    (u_2 - \langle u_2 \rangle_1)^2 \right\rangle_1 
  = \bvcorr-\avcorr .
\end{equation}
The superposition approximation is inconsistent asymptotically in cases
in which $\avsup - \bvsup$ does not decay more rapidly than $\avsup$ as
function of the separation $d$. As remarked in \rcite{ODD05a}, there
are indeed relevant cases in which this consistency condition is not
fulfilled (see Subsec.~\ref{sec:asymptotic}).

\subsection{Effective potential in the intermediate asymptotic regime $r_0 \ll d \ll L$}
\label{sec:asymptotic}

In this subsection we compute $\vmen(d)$ asymptotically in the
intermediate range $r_0 \ll d \ll L$. For this purpose, we have to
make some restricting assumptions which, however, seem to be satisfied
in the experimental setups investigated so far.
First, in view of the discussion in the Introduction concerning the
electrical fields, we assume the proportionality
\begin{equation}
  \label{eq:scalingPim}
  \Pi_m^2 \sim |\Pi_1 \Pi_2| .
\end{equation}
This is valid if the interface stress is quadratic in a field
satisfying linear superposition in the two--particle configuration.
Examples are given below by some specific electrostatic models (see,
c.f.,
Eqs.~(\ref{eq:Pim_ideal}) and (\ref{eq:Pim_debye})).
Second, we assume that the single--colloid pressure $\Pi$ decays far
from the colloid as 
\begin{equation}
  \label{eq:Pidecay}
  \Pi(r) \sim r^{-n} , \qquad n>4.
\end{equation}
In the experimentally relevant case of charged particles at water
interface, it has been established both theoretically \cite{Hurd85}
and experimentally \cite{ABCF02a} that $n=6$. 
(This can be understood easily: the charge of a particle induces a
screening image charge in the water, so that the distant electric
field is dipolar.)
The constraint $n>4$ will allow us to estimate the
integrals appearing in Eqs.~(\ref{eq:Vmen1},~\ref{eq:Vmen2}) by
approximating them by the contribution of the regions near the
colloids. This condition excludes, however, the effect of an external
electric field (in that case the resulting pressure does not have to decay at all)
and the case that the colloidal charge is not perfectly screened so
that the distant electric field corresponds to a monopole, i.e., $n=4$
(this would occur if both fluid phases are dielectric, e.g., air and
insulating oil).

The quantity $\hepsp - 2 \hepsf$ is the net (vertical) force by an
external agent acting on the total system consisting of two colloids
plus the interface (see Appendix~\ref{app:ufunctions}).
This can be, e.g., gravity (if the colloid is large enough for it to
be quantitatively relevant), dispersion forces by a substrate closely
beneath the interface (this effect can be modeled 
similarly as gravity, see Appendix~\ref{app:flot}), or an optical tweezer
pushing the colloid vertically. Since we have assumed previously that
this external force is independent of the positions of the particles,
it is given by the sum of the net forces in the single--particle
configuration (i.e., if they are infinitely far apart from each
other):
\begin{equation}
  \label{eq:additiveF}
  \hepsp - 2\hepsf = 2 (\epsp - \epsf) .
\end{equation}
From the definitions in Eqs.~(\ref{eq:epsfp},~\ref{eq:hpi}) we can
write
\begin{subequations}
\begin{equation}
  \label{eq:new_hepsp}
  \hepsp = 2 (\epsp + \epsm - \varepsilon_{12}) 
\end{equation}
with 
\begin{equation}
  \label{eq:epsm_def}
  \epsm := \frac{1}{2\pi \gamma r_0} \int_{\hsmen} \!\!\!\!\!\! dA \; \Pi_m ,
  \qquad 
  \varepsilon_{12} := \frac{1}{2\pi \gamma r_0} \int_{{S}_1} \!\!\! dA \; \Pi_2 .
\end{equation}
\end{subequations}
For $d\to\infty$, it is clear that
\begin{equation}
  \varepsilon_{12} \sim \frac{1}{d^{n}} .  
\end{equation}
In this limit, we note that $\epsm$ receives its main contribution
from the regions around $S_\alpha$ implying
\begin{equation}
  \label{eq:approx_epsm}
  \epsm \sim \frac{2}{\gamma r_0}   
  \int^\infty_{r_0} \!\!\!\!\!\! dr \; r \, \Pi_m (r)
  \sim \frac{1}{d^{n/2}} 
  \int^\infty_{r_0} \!\!\!\!\!\! dr \; \frac{r}{r^{n/2}} , 
\end{equation}
provided $n>4$, because $|\Pi_m (\br)| \sim \sqrt{\Pi(d) \, \Pi
  (|{\bf r}-{\bf r}_\alpha|)}$ in those regions which provide the
dominant contribution to the integral.
Therefore, from Eqs.~(\ref{eq:hepsf}, \ref{eq:additiveF})
one obtains asymptotically
\begin{equation}
  \label{eq:addF}
  \varepsilon_m = \epsm - \varepsilon_{12} \sim \epsm \sim 
  \frac{1}{d^{n/2}} .
\end{equation}

With these simplifying assumptions, we shall compute analytically the
behavior of $\vmen (d\to\infty)$ to leading order in $1/d$. More
precisely, on dimensional
grounds\footnote{Due to the rapid decay of $\hat{\Pi}$ far from the
  particles, in the limit
  $L\to\infty$ the length scale $L$ enters into the problem only if 
  $\varepsilon_\Pi \neq \varepsilon_F$ and in that case just as an 
  upper bound
  on $d$ to avoid a logarithmic divergence (see the discussion after
  Eq.~(\ref{eq:single_u})).} the expansion parameter is $r_0/d$.
In principle, $\hat{\Pi}$ can contain and thus introduce additional
length scales, for example the Debye length if the fluid phase is an
electrolyte. This complicates the problem, which then has to be
analyzed numerically
(see Subsec.~\ref{sec:debye}). \\

\noindent {\bf The superposition approximation}: Within the
superposition approximation $\vmen(d)$ was computed in detail in
\rcite{ODD05a}. Here we sketch briefly the estimate of the asymptotic
behavior of $\bvsup$ (compare the three terms in
Eq.~(\ref{eq:Vsup2})):
\begin{subequations}
\begin{equation}
  \int_{\hsmen} \!\!\!\!\!\! dA \; \Pi_1 u_2
  \sim
  2\pi\gamma r_0 \epsp u(d) +  
  \Pi(d) \left[ 2 \int_{{S}_{\rm men, 2}} \!\!\!\!\!\! dA \; u_2
    - \frac{1}{2} \pi r_0^3 \epsp + \pi r_0^2 \langle u_2 \rangle_2
  \right] ,
\end{equation}
because the main contribution stems from the regions around
$S_\alpha$, and
\begin{equation}
  \int_{S_1} \!\!\! dA \; \Pi_2 u_2
  \sim
  \Pi(d) u(d) \int_{S_1} \!\!\! dA \; ,
\end{equation}
\begin{equation}
  2\pi\gamma r_0 \epsf \langle u_2 \rangle_1
  \sim 2\pi\gamma r_0 \epsf \left[ u(d) - 
    \frac{1}{4 \gamma} r_0^2 \, \Pi(d) \right] ,
\end{equation}
after expanding around $r_2 = d$. Thus
\begin{align}
  \bvsup(d) &\sim 2\pi\gamma r_0 (\epsf - \epsp) u(d) 
  - \Pi(d) \left[ 
    2 \int_{{S}_{\rm men, 2}} \!\!\!\!\!\! dA \; u_2
    + \pi r_0^2 \langle u_2 \rangle_2 \right] ,
\end{align}
\end{subequations}
because $\Pi(d)$ is asymptotically subdominant compared with $u(d)$ 
(see Eqs.~(\ref{eq:Pidecay}, \ref{eq:single_u})).
The difference $^a V_{\rm sup}-^b V_{\rm sup}$ in
Eq.~(\ref{eq:discrepancy}) between the two implementations of the
superposition approximation can be estimated as \cite{ODD05a} (see
Eq.~(\ref{eq:single_u}))
\begin{subequations}
\label{eq:discr_asymp}
\begin{align}
  \left\langle (u_1 +u_2) \frac{\partial u_2}{\partial n_1} 
  \right\rangle_1 
  &\approx \frac{1}{2} r_0 \langle u_1 \rangle_1 \nabla^2 u(d) 
  \propto \Pi(d) \sim \frac{1}{d^n} , \\
  \left\langle (u_2 - \langle u_2 \rangle_1)^2 \right\rangle_1 
  &\approx \frac{1}{2} r_0^2 |\nabla u(d)|^2 
  \sim \left\{
    \begin{aligned}
      1/d^{2(n-1)} ,
      & \quad \textrm{if } \epsf-\epsp = 0 , \\
      1/d^2 ,
      & \quad \textrm{if } \epsf-\epsp \neq 0 .
    \end{aligned}
  \right.
\end{align}
\end{subequations}
Thus, whenever $\epsf-\epsp \neq 0$ (i.e., the system is not
mechanically isolated in the sense that there must be a force acting
on the boundary $C_L$ of the interface to compensate this non--vanishing net
force) one finds
\begin{equation}
  \label{eq:vsup_asymp}
  \bvsup(d) \sim 2\pi\gamma r_0 (\epsf - \epsp)
  u(d) \sim 2\pi\gamma r_0^2 (\epsf-\epsp)^2 \ln\frac{d}{L} + 
  \rm{const} ,
\end{equation}
corresponding to an attractive force irrespective of the precise form
of the function $\Pi(r)$. 
(The additive constant, which does not affect the physical
conclusions, depends on the precise form of the 
boundary condition at $C_L$.)
Physically, $\bvsup$ is the work done by the net force
$2\pi\gamma(\epsf-\epsp)$ upon a vertical shift of the subsystem 
consisting of one colloid plus its surrounding
interface (behaving like an ``effective particle'') 
by an amount $u(d)$ due to the deformation induced by the
second colloid.  In this case (i.e., $\epsf-\epsp \neq 0$), 
the difference
$\avsup-\bvsup$
decays more rapidly than $\bvsup (d)$ and both expressions
$\avsup$ and $\bvsup$ agree asymptotically.
Equation~(\ref{eq:vsup_asymp}) exhibits the same dependence on $d$ as the
potential energy associated with the flotation force; this is
discussed briefly in Appendix~\ref{app:flot}.

If $\epsf-\epsp=0$ (corresponding to mechanical isolation),
$\bvsup(d\to\infty) \sim 1/d^n$ (see Eq.~(\ref{eq:Pidecay})).  This decay agrees with previous
findings \cite{MeAi03,FoWu04,ODD05a}, but the reliability of this
result is unclear because the difference $\avsup-\bvsup$
decays with the same power law. As a matter of fact, the amplitudes of
the asymptotic decay of $\avsup$ and $\bvsup$ differ and are in turn
different from Eq.~(52) in \rcite{ODD05a}, because there actually
neither of the two representations of ${\cal \hat{F}}_{\rm eq}$ was
used. However, for the leading behavior this is unimportant because
$\vmen$ is asymptotically dominated by the
correction to the superposition approximation. \\

\noindent {\bf Beyond the superposition approximation}: If the
interface deformation field in Eq.~(\ref{eq:hu}) is evaluated near
colloid 1, the term $u_2+u_m$ is dominated by $u_2$ in the absence of
mechanical isolation and by $u_m$ in the case of mechanical isolation 
(see, c.f., Eqs.~(\ref{eq:single_u}, \ref{eq:cross_u})):
\begin{equation}
  u_2 + u_m \sim \left\{
    \begin{aligned}
      u_m(d) \sim d^{-n/2} ,
      & \quad \textrm{if } \epsf-\epsp = 0 , \\
      u_2(d) \sim \ln d ,
      & \quad \textrm{if } \epsf-\epsp \neq 0 . 
    \end{aligned}
    \right.
\end{equation}
Thus the superposition approximation holds if and only if 
the system is {\em
  not} mechanically isolated. Otherwise, the correction $u_m$ is
dominant and, for asymptotically large separations $d$, $\vmen \sim
V_{\rm corr}$. Hence in the following we take $\epsf-\epsp=0$. Since
the main contributions to the integrals stem from the regions around
$S_\alpha$ one obtains the estimates
\begin{equation}
  - \int_{\hsmen} \!\!\!\!\!\! dA \; 
  \left[\Pi_2 \, u_m + 2 \Pi_m \, u_2 + \Pi_m \, u_m \right] 
  \sim - \int_{{S}_{\rm men, 2}} \!\!\!\!\!\! dA \; 
  \left[\Pi_2 \, u_m + 2 \Pi_m \, u_2 \right] \sim \frac{1}{d^{n/2}} .
\end{equation}
Using the asymptotic decay of $u_m$ given by Eq.~(\ref{eq:cross_u}) we
find
\begin{equation}
  \label{eq:bvcorr_asymp}
  \bvcorr(d) \sim 
    - \int_{{S}_{\rm men, 2}} \!\!\!\!\!\! dA \; 
    \left[\Pi_2 \, u_m + 2 \Pi_m \, u_2 \right] 
    + 2 \pi \gamma r_0 \left\langle \epsf u_m + \epsm u_1 
    \right\rangle_1 
    - 4 \pi \gamma r_0^2 \epsf \epsm 
    \sim \frac{1}{d^{n/2}} ,
\end{equation}
because each term scales like $d^{-n/2}$ and there is no reason for
mutual cancellations. In this expression one can identify two distinct
contributions with a simple physical meaning. After inserting
Eqs.~(\ref{eq:hu},~\ref{eq:hepsf}) into the
relationship~(\ref{eq:equil_h}) we write $\Delta\hat{h}_2 = \Delta h_2
+ \Delta h_m$ with
\begin{subequations}
\begin{align}
  \Delta h_2 &:= \langle u_1 + u_2 \rangle_2 - r_0 \, \epsf , 
  \quad (\sim \langle u_2 \rangle_2 - r_0 \, \epsf + {\cal O}(d^{-n})) \\
  \Delta h_m &:= \langle u_m \rangle_2 - r_0 \, \varepsilon_m
  \quad (\sim d^{-n/2}) ,
\end{align}
\end{subequations}
where the asymptotic decays are giving by Eqs.~(\ref{eq:addF},~\ref{eq:cross_u}).
Accordingly, expression~(\ref{eq:bvcorr_asymp}) shows that $\bvcorr$ is
dominated asymptotically by\footnote{Note that by symmetry $\langle
  u_2 \rangle_2 = \langle u_1 \rangle_1$ and $\langle u_m \rangle_2 =
  \langle u_m \rangle_1$.} (i) the work done by the additional pressure
$2\Pi_m$ if the single--colloid configuration is deformed relative to
the reference configuration, that is, $-\int dA \; 2 \Pi_m \,
u_2 + 2\pi \gamma r_0 \epsm \Delta h_2$,
and (ii) the work done by the
forces acting in the single--colloid configuration upon the additional displacement $u_m$, that is, $-\int dA
\; \Pi_2 \, u_m + 2\pi \gamma r_0 \epsf \Delta h_m$.

The asymptotic behavior of $\avcorr(d\to\infty)$ can be derived by
estimating the behavior of the terms in Eq.~(\ref{eq:Vcorr1})
individually, as in the case of Eq.~(\ref{eq:Vcorr2}), with the result
\begin{equation}
  \label{eq:avcorr_asymp}
  \avcorr(d) \sim 
  - 2 \gamma \int_{S_{\rm men, 2}} \!\!\!\!\!\! dA \; 
  (\nabla u_m) \cdot (\nabla u_2)
  - 4 \pi \gamma r_0^2 \epsf \epsm 
  \sim \frac{1}{d^{n/2}} .
\end{equation}
The asymptotic decay predicted by Eqs.~(\ref{eq:bvcorr_asymp}) and
(\ref{eq:avcorr_asymp}) must agree both with respect to the decay
exponents and the amplitudes, because the difference $^a V_{\rm
  sup}-^b V_{\rm sup}$ decays asymptotically more rapidly (see
Eq.~(\ref{eq:discr_asymp}) for $\epsf-\epsp = 0$).
The sign of the force described by $V_{\rm corr}(d)$ is not evident
from the outset, but in the applications we shall consider later, it
turns out to be always attractive.

It is interesting to compare our result with the corresponding one in
\rcite{WuFo05}. After noting the equivalence $u_m \leftrightarrow 2
u_{12}$ in the notations, one finds that Eq.~(7) in \rcite{WuFo05} is
identical with the integral term in Eq.~(\ref{eq:avcorr_asymp}) here.
We obtain an additional term $\propto \epsf \epsm$ because we treat
the contribution by the colloid to the free energy functional in full
detail, while in \rcite{WuFo05} the point--particle approximation
($r_0 \rightarrow 0$) is used from the outset (compare
Eq.~(\ref{eq:freeF1}) here with Eq.~(1) in \rcite{WuFo05}). 
As a consequence, in \rcite{WuFo05} our additional term is lost as a
singularity of the integral term, which is regularized there by introducing
an unknown cutoff length expected to be of the order of $r_0$ (see
also \rcites{FoWu04,ODD05a}).
The analysis of realistic models in Sec.~\ref{sec:appl} will show that
the quantitative contribution of the term $\propto \epsf \epsm$ to the
effective interaction $\vmen$ is actually larger than but
proportional to the contribution from the other term in
Eq.~(\ref{eq:avcorr_asymp}).
\\

To summarize, the capillary--induced effective interaction between two
colloids is given by (compare Eq.~(\ref{eq:Pidecay}) defining $n$)
\begin{equation}
  \label{eq:vmen_asymp}
  \vmen (d) \sim \left\{
    \begin{aligned}
      V_{\rm corr}(d) & \sim d^{-n/2} , \\
      V_{\rm sup}(d) & \sim \ln d ,
    \end{aligned}\qquad
    \begin{aligned}
      & \textrm{if } \epsf-\epsp = 0 , \\
      & \textrm{if } \epsf-\epsp \neq 0 , 
    \end{aligned}
  \right. \qquad (r_0 \ll d) .
\end{equation}

\section{Applications}
\label{sec:appl}

In this section we compute $\vmen(d)$ for $\epsf=\epsp$ (mechanical
isolation) for different realistic models of $\hat{\Pi}$ derived from
the solution of the electrostatic problem within various
approximations. We note that the asymptotic decay of $\vmen(d)$ is the
same as that of the direct electrostatic repulsion, so that these
detailed calculations beyond the asymptotic analysis of
Subsec.~\ref{sec:asymptotic} are necessary in order to be able to 
address this
fine--tuning problem and to determine whether the total force is
asymptotically attractive or repulsive.

\subsection{Ideally conducting fluid phase}
\label{sec:idealconductor}

The simplest model consists of approximating water by an ideal
conductor.  Formally this corresponds to the limit of zero
temperature, so that the osmotic pressure of the mobile charges
accumulated at the interface vanishes and the Debye length
$\kappa^{-1}$ is zero (see, c.f., Eq.~(\ref{eq:debyelength})). In this case the electric field $\hat{\bf E} = \hat{E}
\ez$ is always normal to the interface and the pressure is given by
Maxwell's stress tensor evaluated at the insulating side of the
interface (we use Gaussian units),
\begin{equation}
  \hat{\Pi} = \frac{1}{4\pi} \left. \ez \cdot \left( 
    \epsilon \hat{\bf E} \hat{\bf E} - 
    \frac{1}{2} \epsilon \hat{E}^2 {\bf 1}
    \right) \cdot \ez \right|_{z=0^+} = 
    \frac{\epsilon_1}{8\pi} \hat{E}^2 (z=0^+) ,
\end{equation}
where $\epsilon_1$ is the dielectric constant of the insulating phase
($z>0$).

In the present context the electrostatic problem of a charged sphere partially immersed in a
conducting fluid has been solved numerically and
semi--analytically in \rcite{DaKr06a}. There it has been shown that
the single--colloid pressure field exhibits an integrable divergence
upon approaching the three--phase contact line and that asymptotically
it displays the familiar dipole behavior.  The following approximate
parametrization (Eq.~(1.4) in \rcite{DaKr06a} expressed in terms of
our notation) incorporates these properties and is sufficiently
accurate for our present purposes:
\begin{equation}
  \label{eq:fitPi}
  \Pi(r) = \frac{\gamma \epsf}{r_0} \, b(\mu) 
  \left[ \frac{r}{r_0}-1 \right]^{\mu-1}
  \left[ \frac{r}{r_0} \right]^{-\mu-5} , \qquad 
  b(\mu) := \frac{1}{6}\mu(\mu+1)(\mu+2)(\mu+3) ,
\end{equation}
where $0<\mu<1$ is a fitting parameter the precise value of which 
depends on
the contact angle $\theta$ and the dielectric constant $\epsilon_1$.
This expression is normalized so that $\epsp=\epsf$ and it corresponds
to an exponent $n=6$ independent of the choice for $\mu$. 
Therefore, far from a colloid, the
single--colloid electric field $|{\bf E}|=\sqrt{8\pi\Pi/\epsilon_1}$ is that
of a dipole perpendicular to the interface, the strength of which is
given by
\begin{equation}
  \label{eq:dipole}
  |p| = \lim_{r \to \infty} \epsilon_1 \, r^3 \,
  |{\bf E}(r)| = 
  \epsilon_1 \, r_0^3 \, 
  \sqrt{\frac{8\pi\gamma\epsf b(\mu)}{\epsilon_1 r_0}} .
\end{equation}

In the presence of two colloids, we take $\hat{{\bf E}} \approx {\bf
  E}_1 + {\bf E}_2$. This approximation allows a simplification of the
calculations and should not alter the physical picture significantly.
The approximative character is due to possible violations of the
electrostatic boundary conditions at the surfaces of the colloidal particles: The
additional polarization of colloid 1 induced by ${\bf E}_2$ will
actually lead to an electric field $\hat{{\bf E}} = {\bf E}_1 + {\bf
  E}_2 + \delta {\bf E}$ in the neighborhood of colloid 1, with an
induced electric field $|\delta {\bf E}(\br,d)| \sim \chi(\br) E_2
(d)$, where the ``effective susceptibility'' $\chi(\br)$ is expected to be at most of
order unity. Thus, our conclusions will be qualitatively correct with
a quantitative error of a factor of order unity.
Under these conditions the field $\Pi_m(\br)$ defined by Eq.~(\ref{eq:hpi}) is given
within this approximation by
\begin{equation} 
  \label{eq:Pim_ideal}
  \Pi_m (\br) = \sqrt{\Pi (|\br - \br_1|) \Pi (|\br - \br_2|)} .
\end{equation}
The integrals in Eq.~(\ref{eq:avcorr_asymp}) are computed by using the
expressions in Eqs.~(\ref{eq:single_u}, \ref{eq:cross_u}) so that
in the limit $r_0 \ll d$ one obtains (for details see
Appendix~\ref{app:ideal_calc})
\begin{subequations}
  \label{eq:ideal_int}
\begin{equation}
  \epsm \approx \frac{4 b(\mu)}{\mu+1} \epsf 
  \left(\frac{r_0}{d}\right)^3 
\end{equation}
and
\begin{equation}
  \int_{S_{\rm men, 2}} \!\!\!\!\!\! dA \; 
  (\nabla u_m) \cdot (\nabla u_2) \approx  
  \frac{8\pi b(\mu)}{\mu+1} M(\mu) \, r_0^2 \epsf^2 
  \left(\frac{r_0}{d}\right)^3 ,
\end{equation}
\end{subequations}
with 
\begin{equation}
  \label{eq:Mfunction}
  M(\mu) := \frac{1}{8}(\mu+1) b(\mu) 
  \int_0^1 \!\!\! dv \; v^4
  \, _2F_1(1-\mu,4;5;v) \, 
  _2F_1(\frac{1-\mu}{2},1;2;v) 
\end{equation}
and $_2F_1(\alpha,\beta;\gamma;z)$ is the
hypergeometric function (see Subsec.~9.1 in \rcite{GrRy94}).
We have checked numerically that $M(\mu) \approx \mu/5$ with a maximum
error of $\approx 1.3\%$ within the range $0\leq\mu\leq 1$. The final
result reads
\begin{equation}
  \label{eq:approxVmen}
  \vmen (d) \approx \avcorr(d) \approx 
  - \frac{16\pi b(\mu)}{\mu+1} [1 + M(\mu)] \,
  \gamma r_0^2 \, \epsf^2 
  \left(\frac{r_0}{d}\right)^3 .
\end{equation}
This corresponds to an attractive force. We note that the dominant
contribution to $\vmen (d)$ stems from the term proportional to $\epsm$ in
Eq.~(\ref{eq:avcorr_asymp}), the more so the smaller $\mu$ is.
Smaller values of $\mu$ correspond to an increasing importance of the
electric field near the colloid (see Eq.~(\ref{eq:fitPi})).

$\vmen(d)$ is to be compared with the potential energy due to the
direct electrostatic repulsion $V_{\rm rep}(d)$ of the charged
colloids. The potential energy of one dipole in the field of another
identical dipole is $p^2 /(\epsilon_1 d^3)$ so that $V_{\rm
  rep}(d) \approx p^2/(2 \epsilon_1 d^3)$ for large $d$. (One must
divide by a factor $2$ because within our model no work is done on the
image charge inside the conducting phase forming the dipole.)
Collecting the results, we find that the total interaction energy at
large separations is given by
\begin{equation}
  \label{eq:Vtotal}
  V_{\rm total}(d) = V_{\rm rep}(d) + \vmen(d) \approx
  4 \pi \gamma r_0^2 \, \epsf b(\mu)
  \left(\frac{r_0}{d}\right)^3 \left[ 1
    - \frac{4}{\mu+1} [1+M(\mu)] \epsf \right] .
\end{equation}
Hence we see that the attractive capillary potential is proportional to 
$\epsf^2$ and the direct electrostatic repulsion is proportional to $\epsf$.
Thus in the limit $\epsf \ll 1$, on which our calculations are
based, the electrostatic repulsion is always larger than the capillary
attraction.  The leading--order analysis of this model predicts an
attraction only if the charge of the colloid is large enough so that
$\epsf = {\cal O}(1)$. The critical value above which there is
attraction is given by
\begin{equation}
  \label{eq:critical}
  \varepsilon_{F, \rm crit} (\mu)= \frac{\mu+1}{4[1+M(\mu)]} ,
\end{equation}
which lies in the range $1/4 < \varepsilon_{F, \rm crit} < 5/12$ for
$0<\mu<1$.

\subsection{Finite Debye length}
\label{sec:debye}

Consider now the more general case of an upper insulating phase
(dielectric constant $\epsilon_1$) and a lower electrolytic phase
(dielectric constant $\epsilon_2$, electrolyte concentration $n_0$) at
a finite temperature $T$. The stress tensor acting on the interface is
due to the difference of Maxwell's stress tensor just above and below
the interface, plus an osmotic pressure $\hat{p}_{\rm osm}$ by the
excess of ions concentrated close to the interface:
\begin{equation}
  \hat{\Pi} = \frac{1}{4\pi} \left. \ez \cdot \left( 
      \epsilon \hat{\bf E} \hat{\bf E} - 
      \frac{1}{2} \epsilon \hat{E}^2 {\bf 1}
    \right) \cdot \ez \right|^{z=0^+}_{z=0^-}
  + \hat{p}_{\rm osm} .
\end{equation}
With $\hat{\Phi}(\br)$ denoting the electrostatic potential at the
interface, $\hat{E}_{z,\pm}(\br) := \ez \cdot \hat{\bf
  E}(\br,z=0^\pm)$ as the normal component of the electric field at the
interface (with $\epsilon_1 \hat{E}_{z,+}=\epsilon_2 \hat{E}_{z,-}$)
and $\Ep(\br, z=0)$ as the (continuous) parallel component at the
interface we have
\begin{equation}
  \label{eq:maxwellPi}
  \frac{1}{4\pi} \left. \ez \cdot \left( 
      \epsilon \hat{\bf E} \hat{\bf E} - 
      \frac{1}{2} \epsilon \hat{E}^2 {\bf 1}
    \right) \cdot \ez \right|^{z=0^+}_{z=0^-} = 
    \frac{\epsilon_2-\epsilon_1}{8\pi} \left[
      \frac{\epsilon_1}{\epsilon_2} \Ezp^2 + \Ep^2 
    \right],
\end{equation}
and with $\Delta n(\br)$ as the excess ion concentration at the interface
\begin{equation}
  \label{eq:posm}
  \hat{p}_{\rm osm} \approx k_B T \Delta n = 
  n_0 k_B T \left[ \exp\left(\frac{e\hat{\Phi}}{k_B T}\right) + 
  \exp\left(- \frac{e\hat{\Phi}}{k_B T}\right) -2 \right] ,
\end{equation}
assuming that the electrolyte is dilute and consists of monovalent
ions.

In order to solve the electrostatic problem we introduce two
simplifications: We apply the Debye--H\"uckel approximation for dilute
electrolytes and we approximate the extended colloid by a point charge
$q$ at its center, i.e., we retain only the monopole term of the
source of the field. This 'monopolar' approximation corresponds
formally to the limit of vanishing contact radius, $r_0 \rightarrow
0$, 
and it can be expected to provide the correct field at distances from
the colloid which are large compared to $r_0$. This is in fact
complementary to the limiting case $\kappa r_0 \gg 1$ considered in
the previous subsection.
From Eq.~(\ref{eq:posm}) we find
\begin{equation}
   \hat{p}_{\rm osm} \approx \frac{\epsilon_2}{8\pi} 
   \kappa^2\,\hat{\Phi}^2 ,
\end{equation}
where the screening length is given by
\begin{equation}
  \label{eq:debyelength}
  \kappa^{-1}=\sqrt{\frac{\epsilon_2 k_B T}{8\pi n_0 e^2}} .
\end{equation}
For pure water $\epsilon_2 = 81$ and $n_0 \approx 10^{-7}$ mol/l 
(that is, $\rm{pH}=7$)
lead to a screening length $\approx 1\, \mu$m at room temperature.

\begin{figure}
  \begin{center}
    \epsfig{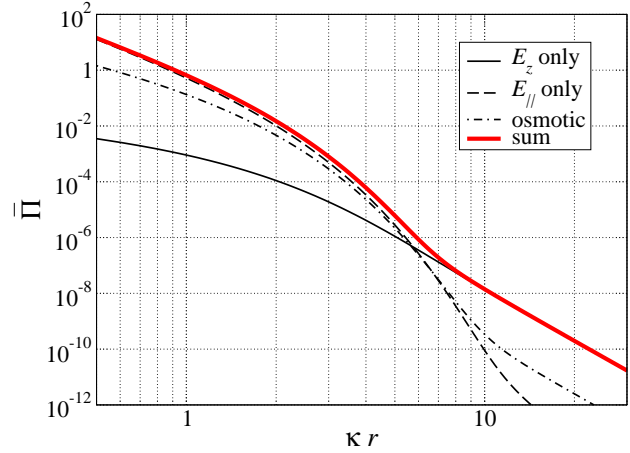}
    \caption{
      The dimensionless stress $\bar\Pi := 
      \Pi/[\gamma r_0 \kappa^2 \epsf/{\cal P}(\kappa r_0)]$
    due to a single point charge at the interface for $\epsilon_2/\epsilon_1=81$, as given by
    Eq.~(\ref{eq:singlepi}) (thick solid line). The thin lines
    represent each of the three additive contributions in
    Eq.~(\ref{eq:singlepi}): due to $E_{z}$, only given by ${\cal I}_b$;
    due to $E_{\parallel}$, only given by ${\cal I}_c$; and 
    due to the osmotic pressure, only given by ${\cal I}_a$.
    As expected, one observes a crossover in $\bar\Pi$ 
    at a distance $r$ comparable with $r_{\rm cross}$ defined by
    Eq.~(\ref{eq:rcross}).}
    \label{fig:cross}
  \end{center}
\end{figure}

We consider first the case of a single colloid corresponding to a
point charge $q$ located at the flat interface between an insulator
and an electrolyte.  Proceeding along the lines of, e.g., \rcites{Hurd85,ACNP00}, we find:
\begin{subequations}
  \label{eq:Efields}
\begin{eqnarray}
  \Phi (r) &=& \frac{2q}{\epsilon_2 r}\; {\cal I}_a(\kappa r) , \\
  E_{z,+}(r) &=& -\frac{2q}{\epsilon_2 r^2}\; 
  {\cal I}_b(\kappa r) , \\
  {\bf E}_{\parallel} (r) = \mbox{} - \frac{d \Phi}{d r} \er &=& 
  \frac{2q}{\epsilon_2 r^2}\; 
  {\cal I}_c(\kappa r) \; \er,
\end{eqnarray}
\end{subequations}
where $\er$ is the unit radial vector pointing away from the center of the
colloid, and the auxiliary functions ${\cal I}_n(k)$ are given by
integrals over a Bessel function:
\begin{subequations}
\label{eq:Idef}
\begin{eqnarray}
  {\cal I}_a(k) &:=& \int_0^\infty dx\; J_0(x)\;
  \frac{x}{\frac{\epsilon_1}{\epsilon_2} x +
    \sqrt{x^2+k^2}} , \\
  {\cal I}_b(k) &:=& \frac{\epsilon_2}{\epsilon_1} \int_0^\infty dx\; J_0(x)\;
  \frac{x\,\sqrt{x^2+k^2}}{
    \frac{\epsilon_1}{\epsilon_2} x + \sqrt{x^2+k^2}} , \\
  {\cal I}_c(k) &:=& {\cal I}_a(k) - k\, \frac{d\, {\cal I}_a}{d k} (k).
\end{eqnarray}
\end{subequations}
These integrals have been computed numerically; the analytic
expressions for their asymptotic behaviors are derived in 
Appendix~\ref{app:Ifunctions}. Thus, taking $\epsilon_2/\epsilon_1 \gg 1$ (the
ratio $\epsilon_2/\epsilon_1$ is approximately $81$ for
water in contact with air) 
in the intermediate asymptotic regime $1 \ll \kappa r \ll
(\epsilon_2/\epsilon_1)^2$ one obtains the following expressions for the potential
and the field components (see Eq.~(\ref{eq:Efields})):
\begin{subequations}
  \label{eq:asymp}
\begin{eqnarray}
  \Phi(r) & \approx & \frac{2q}{\epsilon_2 r} \left[
    \frac{\epsilon_1}{\epsilon_2} \frac{1}{(\kappa r)^2}
    + {\rm e}^{-\kappa r} \right] , \\
  E_{z,+}(r) & \approx & \frac{2q}{\epsilon_2 r^2} \left[
    - \frac{1}{\kappa r} + \sqrt{\frac{\pi}{2}} \;
    \frac{\epsilon_1}{\epsilon_2} \; (\kappa r)^{3/2}\, 
    {\rm e}^{-\kappa r} \right] , \\
  {\bf E}_{\parallel}(r) & \approx & \frac{2q}{\epsilon_2 r^2} \left[
    3 \frac{\epsilon_1}{\epsilon_2} \frac{1}{(\kappa r)^2}
    + \kappa r \, {\rm e}^{-\kappa r} \right] \er .
\end{eqnarray}
\end{subequations}
Asymptotically $\Phi(r)$ decays as $1/r^3$ and the electrostatic
interaction energy $q \Phi (d)$ of a second charge $q$ at a distance
$d$ from the first one decays likewise as $1/d^3$. This is the
celebrated dipole repulsion between charged colloidal particles first
conjectured in \rcite{Pier80}. However, due to the large value of the
ratio $\epsilon_2/\epsilon_1$, closer to the charge there is a crossover to a screened
Coulomb potential $\Phi(r) \propto \exp{(-\kappa r)}/r$. One can introduce a crossover length $r_{\rm cross}$
defined from the asymptotic behavior given by Eq.~(\ref{eq:asymp}) as
\begin{equation}
  \label{eq:rcross}
  \frac{\epsilon_1}{\epsilon_2} \frac{1}{(\kappa r_{\rm cross})^2}
  = {\rm e}^{-\kappa r_{\rm cross}} .
\end{equation}
This equation has solutions only if $\epsilon_1/\epsilon_2 < (2/{\rm
  e})^2 \approx 0.54$, in which case the relevant solution $\kappa
r_{\rm cross}$ is larger than $2$ and depends only weakly, i.e., logarithmically on the precise value of the ratio
$\epsilon_1/\epsilon_2$. 
For $\epsilon_2/\epsilon_1=81$ this gives
$\kappa r_{\rm cross} \approx 8.7$. The parallel component
${\bf E}_{\parallel}$ exhibits the same crossover, but the normal component
$E_{z,+}$ always decays algebraically and is much larger than the
parallel component.
Nevertheless, the contribution of $E_{z,+}$ to the pressure field
$\Pi(r)$ is reduced by a factor $\epsilon_1/\epsilon_2$ (see
Eq.~(\ref{eq:maxwellPi})), so that $\Pi(r)$ will also exhibit the
crossover at a distance $r \approx r_{\rm cross}$. From Eqs.~(\ref{eq:maxwellPi}, \ref{eq:posm}) one finds
\begin{equation}
  \label{eq:singlepi}
  \Pi(r) = \gamma r_0 \kappa^2 
  \frac{\epsf}{(\kappa r)^4 {\cal P} (\kappa r_0)} \left[ 
    \frac{\epsilon_1}{\epsilon_2}
    \left(1-\frac{\epsilon_1}{\epsilon_2}\right)
    {\cal I}_b^2(\kappa r)
    + \left(1-\frac{\epsilon_1}{\epsilon_2}\right) 
    {\cal I}_c^2(\kappa r)
    + (\kappa r)^2 {\cal I}_a^2(\kappa r)
  \right] ,
\end{equation}
where (see Eq.~(\ref{eq:epsfp})) 
\begin{equation}
  \label{eq:qeff}
  \epsf = \epsp = \frac{q^2 \kappa^2}{2 \pi \epsilon_2 \gamma r_0}
  {\cal P}(\kappa r_0)
\end{equation} 
with the dimensionless function
\begin{eqnarray}
  \label{eq:calP}
  {\cal P}(\kappa r_0) & := &
  \int_{\kappa r_0}^{\infty} \!\! dx \;
  \frac{1}{x^3} \left[ 
    \frac{\epsilon_1}{\epsilon_2}
    \left(1-\frac{\epsilon_1}{\epsilon_2}\right)
    {\cal I}_b^2(x)
    + \left(1-\frac{\epsilon_1}{\epsilon_2}\right) 
    {\cal I}_c^2(x) + x^2 {\cal I}_a^2(x)
  \right] \\
  & & \nonumber \\
  & \approx & 
  \left[\frac{1}{\kappa r_0} + \frac{1}{2 (\kappa r_0)^2} \right]
  {\rm e}^{-2\,\kappa r_0} , \nonumber
\end{eqnarray}
where the last line is the leading (zeroth order) contribution in an expansion in terms of the small parameter
$\epsilon_1/\epsilon_2$.
Equation~(\ref{eq:qeff}) allows one to express the 
parameter $q$, the value of which is often uncertain, 
in terms of the more convenient parameter $\epsf$.
Figure~\ref{fig:cross} shows how $\Pi(r)$ attains the dipole limiting
behavior $\propto 1/r^6$ due to the electrostatic pressure $\propto
E_{z,+}^2$ beyond the crossover distance; at shorter distances,
$\Pi(r)$ is dominated by the term $\propto {\bf E}_{\parallel}^2$ and, 
to a lesser extent, by the osmotic pressure.
The figure indicates also that the main contribution to the total
force $\epsp$ stems from the regions close to the particle, as
evidenced also by the formal divergence of ${\cal P}(\kappa r_0 \to 0)$ 
(see Eq.~(\ref{eq:calP})). Thus, the precise value of $\epsp$ will be
affected by the fact that
our solution of the electrostatic problem within the monopolar
approximation is expected to be reliable in principle only sufficiently
far from the particle.

\begin{figure}[t]
  \begin{center}
    \vspace*{0.5cm}
    \epsfig{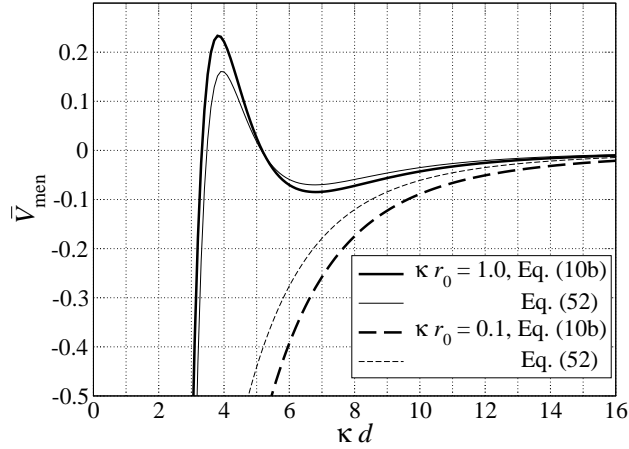}
    \caption{The dimensionless capillary--induced potential energy
      $\bar{V}_{\rm men} := 10^2 \vmen/ [\gamma r_0^2 \epsf^2/ {\cal P}
      (\kappa r_0)]$ for two different values of $\kappa r_0$ and the
      choice $\epsilon_2/\epsilon_1=81$. Thick lines correspond to the
      capillary--induced potential given by Eq.~(\ref{eq:Vcorr2}) 
      (see Eq.~(\ref{eq:vmen_asymp})), whereas thin
      lines show the approximation $\propto \epsf \epsm$ given by
      Eq.~(\ref{eq:vmen_debye}).
      If $\kappa r_0$ is large enough, a minimum appears. 
    }
    \label{fig:vmen_scaling}
  \end{center}
\end{figure}

We consider now two identical point charges at the interface separated
by a distance $d$. Within the Debye--H\"uckel and point--charge
approximations, the solution of this electrostatic problem is given by
the superposition of the single--colloid fields.
In this case the field $\Pi_m(\br)$ in Eq.~(\ref{eq:hpi}) reads (as in
Subsec.~\ref{sec:exact}, the subscript $\alpha(=1,2)$ denotes that the
corresponding field is evaluated with respect to
particle $\alpha$;
see Fig.~\ref{fig:2coll_top} for the notation):
\begin{eqnarray}
  \label{eq:Pim_debye}
  \Pi_m(r) & = &
  \frac{\epsilon_2 - \epsilon_1}{8\pi} \left[ 
    \frac{\epsilon_1}{\epsilon_2} (E_{z,+})_1 (E_{z,+})_2
    + ({\bf E}_{\parallel})_1 \cdot ({\bf E}_{\parallel})_2 \right] 
  + \frac{\epsilon_2}{8\pi} \kappa^2 \Phi_1 \Phi_2 \nonumber \\
  & & \nonumber \\
  & = & 
  \frac{q^2 \kappa^4}{2 \pi \epsilon_2} 
  \frac{1}{\kappa^4 (|\br -\br_1| |\br -\br_2|)^2} \left[ 
    \frac{\epsilon_1}{\epsilon_2}
    \left(1-\frac{\epsilon_1}{\epsilon_2}\right)
    {\cal I}_b(\kappa |\br -\br_1|) {\cal I}_b(\kappa |\br -\br_2|)
    + \mbox{} \right. \nonumber \\
  & & \nonumber \\
  & & \left(1-\frac{\epsilon_1}{\epsilon_2}\right) 
    {\cal I}_c(\kappa |\br -\br_1|) {\cal I}_c(\kappa |\br -\br_2|)
    \frac{\br -\br_1}{|\br -\br_1|} \cdot 
    \frac{\br -\br_2}{|\br -\br_2|} + \mbox{} \nonumber \\
  & & \nonumber \\
  & & \left. \frac{}{} \kappa^2 |\br -\br_1| |\br -\br_2| \,
    {\cal I}_a(\kappa |\br -\br_1|) {\cal I}_a(\kappa |\br -\br_2|)
  \right] .
\end{eqnarray}
Therefore one has (see Eqs.~(\ref{eq:epsm_def}, \ref{eq:qeff}))
\begin{equation}
  \epsm = 
  \frac{{\cal P}_m(\kappa r_0, \kappa d)}{{\cal P}(\kappa r_0)}
  \, \epsf ,
\end{equation} 
where
\begin{eqnarray}
  \label{eq:defPm}
  {\cal P}_m (\kappa r_0, \kappa d)& := & \frac{1}{2\pi} 
  \int_{\hat{S}_{\rm men}} \!\!\!\!\!\!\!\! d^2 {\bf x} \,
  \frac{1}{(x_1 x_2)^2} 
  \left[ \frac{\epsilon_1}{\epsilon_2}
    \left(1-\frac{\epsilon_1}{\epsilon_2}\right)
    {\cal I}_b(x_1) {\cal I}_b(x_2) + \mbox{} \right. \\ 
  & & \nonumber \\
  & & \left. \left(1-\frac{\epsilon_1}{\epsilon_2}\right) 
    {\cal I}_c(x_1) {\cal I}_c(x_2)
    \, \frac{{\bf x}_1 \cdot {\bf x}_2}{x_1 x_2}
    + x_1 x_2 \, 
    {\cal I}_a(x_1) {\cal I}_a(x_2)
  \right] \nonumber
\end{eqnarray}
in terms of ${\bf x}:=\kappa \br$ and ${\bf x}_\alpha := \kappa (\br -
\br_\alpha)$. Dependences on $r_0$ and $d$ enter through the
specification of the integration domain $\hat{S}_{\rm men}$.
Using this expression and Eq.~(\ref{eq:Idef}), we have integrated
numerically Eq.~(\ref{eq:cross}) using bipolar coordinates in a domain
corresponding to a size $L \approx 100 / \kappa$.  Taking this
numerical solution and the exact expression in
Eq.~(\ref{eq:single_u}), we have subsequently computed numerically the
effective interaction energy as $\vmen(d)\approx \bvcorr(d)$ by using
Eq.~(\ref{eq:Vcorr2}). As in the case studied in
Subsec.~\ref{sec:idealconductor} (see the text after
Eq.~(\ref{eq:approxVmen})), the last term in Eq.~(\ref{eq:Vcorr2})
turns out to be larger in magnitude than the remaining terms but
they are approximately proportional to each other, so that for the
purpose of understanding the numerical results we can consider the
following proportionality (upon the application of Eq.~(\ref{eq:addF}))
\begin{equation}
  \label{eq:vmen_debye}
  \vmen(d) \propto 
  \mbox{}- 2 \pi \gamma r_0^2 \varepsilon_m ( 2\epsf + \varepsilon_m)
  \sim - 4\pi \gamma r_0^2 \epsf \epsm =
  - 4\pi \gamma r_0^2 \epsf^2 \,
  \frac{{\cal P}_m(\kappa r_0, \kappa d)}{{\cal P}(\kappa r_0)} .
\end{equation}
Figure~\ref{fig:vmen_scaling} shows indeed that for two choices of $\kappa r_0$ this approximation is
reasonable at large $d$. Numerically we find that the
term $\propto \epsf \epsm$ contributes ca.~$70\%$ of the total
meniscus--induced potential $\vmen$.
Asymptotically for large $d$ it exhibits the predicted
$1/d^3$--behavior (Eq.~(\ref{eq:vmen_asymp}) with $n=6$). More
precisely, the asymptotic behavior of $V_{\rm men}$ can be obtained as
an expansion in terms of $1/d$ by utilizing the two--peak structure of
$\Pi_{\rm m}$ (as for the general discussion in
Subsec.~\ref{sec:asymptotic}): One assumes that for $d\to\infty$ 
the main contribution
to the integral in Eq.~(\ref{eq:defPm}) stems from the
regions near the colloidal particles. The leading asymptotic behavior
of ${\cal P}_m$ is given simply by the lowest--order term of a
Taylor--expansion about $x_1 \approx \kappa r_0, x_2 \approx \kappa d$
and $x_1 \approx \kappa d, x_2 \approx \kappa r_0$ in the integral in
Eq.~(\ref{eq:defPm}) with the asymptotic decay 
for the ${\cal I}$--functions obtained in
Appendix~\ref{app:Ifunctions}\footnote{Next-to-leading terms in the expansion 
  require a more
  elaborate calculation than the simple Taylor--expansion, which
  yields undefined expressions for them.}:
\begin{equation}
 \label{eq:pmasy}
  {\cal P}_m (\kappa r_0, \kappa d) \approx 
  \frac{\epsilon_1}{\epsilon_2}\frac{p_m(\kappa r_0)}{(\kappa d)^3} , 
\end{equation}
where
\begin{eqnarray}
  p_m (\kappa r_0) & := & 2 \int_{\kappa r_0}^\infty dx\,
  \left[ 
    \left(1-\frac{\epsilon_1}{\epsilon_2}\right)
    \frac{{\cal I}_b(x)}{x} + {\cal I}_a(x)
  \right] \\
  & & \nonumber \\
  & \approx & 2 \left[
    \kappa r_0 \, I_0\left(\frac{\kappa r_0}{2}\right) \,
    K_1\left(\frac{\kappa r_0}{2}\right)
    - 1 + {\rm e}^{-\kappa r_0}
    \right] , \nonumber
\end{eqnarray}
and the second line is the leading (zeroth order) contribution in an expansion in terms of the small
parameter $\epsilon_1/\epsilon_2$, obtained from applying
Eqs.~(\ref{eq:Iaeps}, \ref{eq:Ibeps}). According to this expression,
$p_m (\kappa r_0 \to 0) = 4$ and $p(\kappa r_0)$ decreases
monotonously with an asymptotic decay $p(\kappa r_0 \to\infty) \sim
2/(\kappa r_0)$.
This result for ${\cal P}_m (\kappa r_0, \kappa d)$ means that the
asymptotic $1/d^3$-decay is determined by the normal component
$E_{z}(d)$ of the field and
by the potential $\Phi(d)$ stemming from the osmotic pressure. There
is no contribution from ${\bf E}_\parallel$ to leading order in $1/d$
due to the geometrical factor ${\bf x}_1 \cdot {\bf x}_2$ appearing in
Eq.~(\ref{eq:defPm}).
Finally, from Fig.~\ref{fig:vmen_scaling} we infer that if $\kappa
r_0$ is large enough, the asymptotic decay breaks down as a reliable
approximation and a minimum appears in $\vmen(d)$ at a separation $d$ 
which is a few times $\kappa^{-1}$.

\begin{figure}[t]
  \begin{center}
    \vspace*{0.5cm}
    \epsfig{file=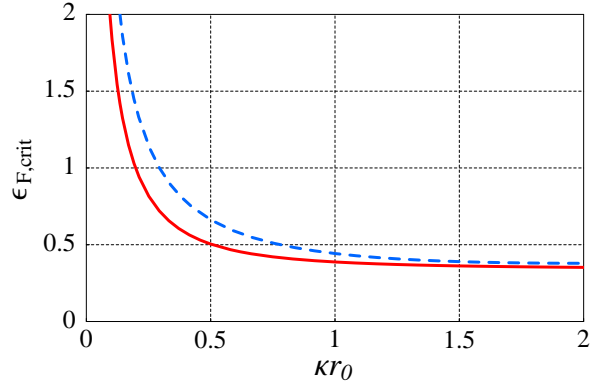,width=.5\textwidth}
    \caption{The solid line gives the critical value $\varepsilon_{F,
        \rm crit}$ as a function of $\kappa r_0$ specified by the fit
      to the numerical results (Eq.~(\ref{eq:fit_critF})). The dashed
      line corresponds to the approximate expression provided by
      Eq.~(\ref{eq:approx_critF}).}
    \label{fig:critF}
  \end{center}
\end{figure}

The energy due to the direct repulsion of the two colloids is given by
Eq.~(\ref{eq:Efields}a):
\begin{equation}
 V_{\rm rep}(d) = q \Phi(d) = 
 4\pi \gamma r_0^2 \epsf 
 \frac{{\cal I}_a(\kappa d)}{(\kappa d) (\kappa r_0) 
   {\cal P}(\kappa r_0)} ,
\end{equation}
and the total energy is $V_{\rm tot}(d) = V_{\rm rep}(d) + \vmen(d)$.
With the approximation~(\ref{eq:vmen_debye}) one has
\begin{equation}
  \label{eq:vtot_debye}
  V_{\rm tot}(d) \approx 
  \frac{4\pi \gamma r_0^2 \epsf}{\kappa r_0\,{\cal P}(\kappa r_0)}
  \left[
    \frac{{\cal I}_a(\kappa d)}{\kappa d} 
    - \epsf \, \kappa r_0 \, {\cal P}_m(\kappa r_0,\kappa d) 
  \right] .
\end{equation}
Asymptotically for $\kappa d \gg 1$ this expression reduces to (see
Eqs.~(\ref{eq:pmasy}, \ref{eq:iaasy}))
\begin{equation}
  \label{eq:Vkappa_asymp}
  V_{\rm tot}(d) \sim 
  4\pi \gamma r_0^2 \epsf 
  \frac{\epsilon_1/\epsilon_2}{(\kappa r_0)^4\,{\cal P}(\kappa r_0)}
  \left(\frac{r_0}{d}\right)^3
  \left[ 1
    + \frac{\epsilon_2}{\epsilon_1} (\kappa d)^2 \, 
    {\rm e}^{-\kappa d}
    - \kappa r_0 \, p_m(\kappa r_0) \, \epsf 
  \right] ,
\end{equation}
to be compared with the potential obtained in the previous subsection~(Eq.~(\ref{eq:Vtotal})). As in that case, the capillary--induced
potential $\vmen$ is reduced by a factor $\epsf \ll 1$ with respect to
$V_{\rm rep}$, and the total potential can be asymptotically
attractive only if $\epsf$ is above a critical value 
\begin{equation}
  \label{eq:approx_critF}
  \varepsilon_{F,\rm{crit}} = \frac{1}{\kappa r_0 \, p_m(\kappa r_0)} ,
\end{equation}
which to leading order is independent of the small ratio 
$\epsilon_1/\epsilon_2$ and turns out to be bounded as
$\varepsilon_{F,\rm{crit}} > 0.38$ (see Fig.~\ref{fig:critF}).
Moreover, we note that even if $\epsf \sim 1$ the term in brackets in
Eq.~(\ref{eq:Vkappa_asymp}) changes sign at a separation $d$ 
comparable with the crossover length $r_{\rm cross}$ (see
Eq.~(\ref{eq:rcross})). This is a consequence of the crossover in
$V_{\rm rep}(d)\propto \Phi(d)$ and suggests that although the
potential is asymptotically attractive, it reaches a minimum at a
distance $d\approx r_{\rm cross}$ and turns repulsive for closer
separations. 
This effect can only be enhanced
by the deviation from the asymptotic $1/d^3$--decay in $V_{\rm
  men}(d)$ observed in Fig.~\ref{fig:vmen_scaling} for $\kappa r_0
\sim 1$.

\begin{figure}[t]
  \begin{center}
    \epsfig{file=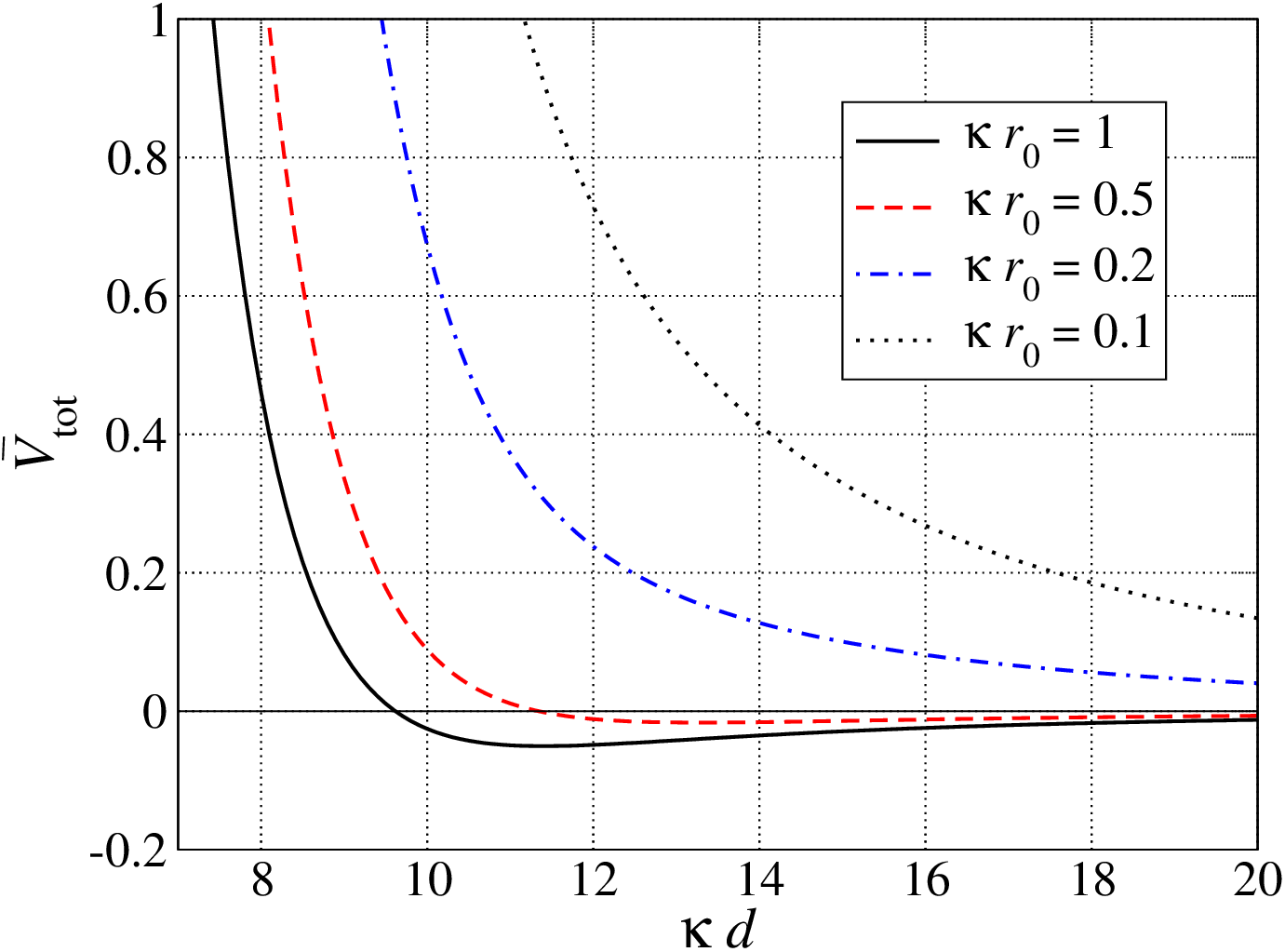,width=.5\textwidth}
    \caption{ The dimensionless total potential energy $\bar{V}_{\rm
        tot} := 10^3 V_{\rm tot} / [\gamma r_0^2 \epsf / {\cal P}
      (\kappa r_0)]$ 
      for different values of $\kappa r_0$ and the choices
      $\epsilon_2/\epsilon_1=81$ and $\epsf=0.6$. Note that the total
      energy is about 10 times smaller than $\vmen$ 
      (the normalization of $\bar{V}_{\rm tot}$ here differs from 
      that of $\bar{V}_{\rm men}$ in Fig.~\ref{fig:vmen_scaling} 
      by a factor $10/\epsf$)
      and exhibits a
      shallow minimum provided that $\kappa r_0$ is not too small.}
    \label{fig:vtot_debye}
  \end{center}
\end{figure}

These conclusions based on the approximation~(\ref{eq:vmen_debye}) are
supported by the corresponding full numerical calculations.
More precisely, from a fit to these numerical results
in the
range $0.1 \leq \kappa r_0 \leq 2.0$ for the ratio
$\epsilon_2/\epsilon_1 = 81$ we find that the critical value is given
approximately by
\begin{equation}
  \label{eq:fit_critF}
  \varepsilon_{F, \rm crit} (\kappa r_0) \approx 
  0.34 \sqrt{1+\frac{0.30}{(\kappa r_0)^2}} \, .
\end{equation}
Figure~\ref{fig:vtot_debye} shows a plot of $V_{\rm tot}(d)$ for a
typical value $\epsf =0.6$, exhibiting a shallow minimum 
if it is asymptotically attractive.
In summary, the capillary--induced attraction can dominate asymptotically
only if $\kappa r_0$ is sufficiently large so that
$\epsf>\varepsilon_{F,{\rm crit}}$, in which case one necessarily has 
$\epsf \gtrsim 1$.

\section{Summary and Discussion}
\label{sec:end}

We have carried out a thorough analysis of the effective interaction
potential $\vmen(d)$ between two colloidal particles floating a
distance $d$ apart from each other at a fluid interface due to the deformation of the
interface caused by the particles. The main result is summarized in
Eq.~(\ref{eq:vmen_asymp}) for the asymptotic behavior of $\vmen(d)$ as
$d\to\infty$. One finds two qualitatively different cases depending on
the (dimensionless) force $\epsf$ acting on the particle, and the 
one, $\epsp$,
acting on the interface (see Eqs.~(\ref{eq:epsfp}, 
\ref{eq:additiveF})): (a) if the system is not mechanically isolated
($\epsf-\epsp \neq 0$), the superposition approximation is valid and
the asymptotic dependence $\vmen(d\to\infty)$ is universal in the
sense of being independent of the detailed distribution of the forces
which deform the interface.  The physical reason for this is that each
particle together with its surrounding interface can be considered as
an effective particle.
(b) In the opposite case of mechanical isolation ($\epsf-\epsp = 0$),
the corrections to the superposition ansatz are dominant and the
asymptotic decay of $\vmen(d)$ depends on the asymptotic properties of
the pressure field acting on the interface. We identified two sources
of violation of the superposition approximation: the boundary
conditions of the deformation at the particle--interface contact line
are violated, and the pressure field acting on the interface does not
satisfy a superposition principle.

These results hold under the rather general assumptions that (i) the
interfacial deformation is small ($\epsf, \epsp \ll 1$),
(ii) the pressure field $\hat{\Pi} = \Pi_1 + \Pi_2 + 2 \Pi_m$ exerted by the
particles on the interface satisfies the scaling relations in
Eqs.~(\ref{eq:scalingPim}, \ref{eq:Pidecay}), and (iii) the external
force acting on the particles is additive (Eq.~(\ref{eq:additiveF})).
Assumptions (i) and (iii) are quite general and there is no evidence
that experimentally they are not fulfilled. Assumption (ii) can be viewed as
a consequence of the condition that the pressure $\hat{\Pi}$ is
derivable from the stress tensor of a Lagrangian quadratic in an
underlying field, as it is typically the case for physical systems:
Besides stresses induced by electric fields as addressed here, there can be, e.g.,
elastic stresses which arise if one of the fluids is in a nematic phase
\cite{ODTD07}. (This latter case provides also an example in which the asymptotic
decay~(\ref{eq:vmen_asymp}) is actually modified by geometrical
constraints which complement the scaling
relation~(\ref{eq:scalingPim}).)

As an application for the case of mechanical isolation, we have
considered the paradigmatic system of electrically charged colloids
(see Figs.~\ref{fig:2coll_side} and~\ref{fig:2coll_top}), which is also of direct experimental
importance. The pressure deforming the interface is due to the
electrostatic field emanating from the particles and gives rise to an
effective attractive interaction $\vmen(d\to\infty) \sim 1/d^3$.
Since the direct electric repulsion between the colloids also decays
$V_{\rm rep} \sim 1/d^3$, this asymptotic analysis is insufficient to
determine whether the total potential $\vmen + V_{\rm rep}$ describes
an asymptotically attractive force as seemingly observed
experimentally. To this end, as a model for the experiments it is necessary to consider in detail the
challenging electrostatic problem posed by two charged spheres
partially immersed in an electrolyte.
Within this approach we have studied the electrostatic problem in two
simplifying limiting regimes: (A) water as one of the fluid phases is
a perfect conductor, and (B) the colloidal particles are replaced by
point charges (monopolar approximation).
Within both approximations we concluded that $\vmen \propto \epsf^2$,
while $V_{\rm rep} \propto \epsf$, so that the direct repulsion
dominates asymptotically; the capillary--induced attraction is
predicted to dominate only if $\epsf$ is larger than a threshold value
$\varepsilon_{F,{\rm crit}} \approx 1$ (see Eqs.~(\ref{eq:critical}, \ref{eq:fit_critF})), i.e., the deformation of the
interface has to be large,
which is outside the range of validity of the present analysis based on small
deformations ($\epsf\ll 1$).

Assumption (A) is a simple model of the experimentally relevant case
of polystyrene or glass colloids floating at the interface between
salty water and air or oil such that the Debye length $\kappa^{-1}$
of water is much smaller than the radius of the particle--interface
contact line $r_0$. Only the residual charges at the interface between
the colloid and the air or oil phase determine $\epsf$. The charge
density $\sigma$ at this interface is rather small compared with the nominal surface density of dissociable groups; we note that in
the case of polystyrene colloids the precise microscopic origin of
this charge density is still unknown.  By dimensional analysis, $\epsf
= (\sigma^2\,r_0/\epsilon_1 \gamma) \, G(\epsilon_{\rm
  C}/\epsilon_1,\theta)$, where $G$ is a dimensionless function
depending on the contact angle $\theta$ and the ratio of dielectric
constants of the colloidal particle and the insulating fluid, 
$\epsilon_{\rm C}$ and $\epsilon_1$, respectively. The
electrostatic solution \cite{DaKr06a} yields $G(1,\theta) \approx
3/\sin^3\theta$ for hydrophobic colloids.
In \rcite{DKB06}\footnote{This work is {\it de facto} a correction
  \cite{ODD06} of previous work \cite{DKB04}.} the single--colloid
meniscus deformation was measured around rather large glass spheres
($r_0 \approx 200\, \mu$m) at a water--oil interface ($\gamma \approx
0.05$ N/m, $\epsilon_1\approx 2$), which are slightly hydrophobic
($\theta \approx 120^o$). Using Eq.~(\ref{eq:single_u}) a value
$\epsf \sim 0.4$ was inferred from the measurements, corresponding to
a charge density of $ \sigma \sim 70\, \mu$C/m$^2$ (or $5\times
10^{-4}$ $e$/nm$^2$) according to the simple
formula given above. This value is close to $\varepsilon_{F,{\rm
    crit}}$ given in Eq.~(\ref{eq:critical}) and thus one could expect
a strongly reduced repulsion or even a net attraction between pairs of
these glass spheres. A corresponding extension of this single--sphere experiment
would be highly desirable.

For truly nanoscopic colloids ($r_0 \alt 1\,\mu$m) with this same
charge density on the air or oil side, the formula above predicts
$\epsf =O(10^{-3})$ and therefore the electrostatic repulsion would
always dominate the capillary attraction. This is consistent with the
results in \rcite{ABCF02a} obtained for polystyrene spheres
($r_0\approx 1\,\mu$m) at the oil/water interface. For highly salty
water, the charge density on the colloid/oil interface is estimated
experimentally to be $\sigma \sim 20\, \mu$C/m$^2$ and force
measurements between two spheres confirmed the repulsive
dipole--dipole interaction with no sign of capillary attraction.

Assumption (B) amounts to modelling a system in which the Debye length of water is comparable
or larger than the radius $r_0$. We have studied in detail the case
that the dielectric constant of one fluid phase (e.g., water) is much
larger than the dielectric constant of the other fluid phase (e.g.,
air). 
The capillary--induced potential $\vmen$ is asymptotically attractive
(see Fig.~\ref{fig:vmen_scaling}) but the total potential $\vmen +
V_{\rm rep}$ can be asymptotically attractive 
for values $\epsf \sim 1$ only if
$\kappa r_0 \gtrsim 1$ (see Fig.~\ref{fig:critF}). However, unlike
the model corresponding to assumption (A), even in such a case 
$\vmen + V_{\rm rep}$ becomes repulsive at small separations
(Fig.~\ref{fig:vtot_debye}). This effect 
can be traced back to a crossover in the interfacial stress
$\Pi (r)$ from the asymptotic algebraic decay (as in model (A)) to an
exponential decay at closer distances $r \lesssim 7/\kappa$ from the
particle (see Fig.~\ref{fig:cross}).

\begin{figure}[t]
  \begin{center}
    \epsfig{file=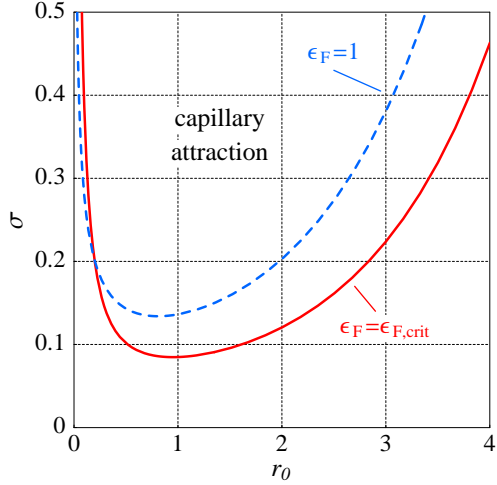,width=.4\textwidth}
    \caption{Parameter space spanned by the surface charge density $\sigma$
      (in units of $e$/nm$^2$) and the contact line radius $r_0$ (in
      $\mu$m) (see text for the fixed values of the other parameters). The solid line corresponds to the loci
      $\epsf=\varepsilon_{F,{\rm crit}}$, so that capillary attraction
      is predicted to occur in systems the parameters of which fall into the region
      above this curve. As a reference curve, the dashed line corresponds to the loci
      $\epsf=1$.}
    \label{fig:modelB}
  \end{center}
\end{figure}

In contrast to the experiments carried out with nano-- and
microcolloids at interfaces with salty water, 
some experiments \cite{QMMH01,GEFM01,NBHD02,TAKK03,GoRu05}
have been performed with microcolloids at interfaces of ultrapure water
such that $\kappa^{-1} \sim r_0 \sim 1$ $\mu$m.
These experiments could have explored phenomena beyond the small deformation regime.
Equation~(\ref{eq:qeff}) provides a relationship between $\epsf$ and
the relevant parameters of the experimental system; regretably, the value of the total charge $q$ is
usually uncertain. In terms of the surface charge density $\sigma$ one
has $q=2\pi\sigma (r_0 \sin\theta)^2 (1+\cos\theta)$, because for ultrapure water the
electrostatic field is dominated by the unscreened charge of the
particle on the water side. For typical values $\epsilon_2=81$,
$\gamma = 0.05\,$N/m, $\kappa = 1\,\mu$m$^{-1}$, 
and $\theta=\pi/2$ (so that $r_0=$ radius of the particle),
Eq.~(\ref{eq:qeff}) gives $\epsf \approx 371 \, \sigma^2 \, r_0^3 \,
{\cal P}(r_0)$ with $r_0$ in $\mu$m and $\sigma$ in units of
$e$/nm$^2$. The values for $\sigma$ quoted in the
literature range from $0.07\,e/$nm$^2$ (\rcite{GhEa97}) to
$0.53\,e/$nm$^2$ (\rcite{SDJ00}). Accordingly, Fig.~(\ref{fig:modelB}) shows
that it seems possible to have capillary attraction (i.e., $\epsf
> \varepsilon_{F,{\rm crit}} \approx 1$) for typical values of the particle radius in the
micrometer range. However, we emphasize that, unlike the conclusion
concerning the asymptotic decay of $\vmen$, the expressions relating the value
of $\epsf$ and $\varepsilon_{F,{\rm crit}}$ with the parameters of the
system involve the behavior of the electric field near the particle
(see Eqs.~(\ref{eq:calP}) and (\ref{eq:pmasy})). Therefore, they are
expected to be affected by corrections to the monopolar approximation
and the Debye--H{\"u}ckel approximation \cite{FDO07}. Thus, further theoretical
work is required to understand these experiments properly and to
arrive at reliable predictions.

\acknowledgments

A.D. acknowledges financial support from the Junta de Andaluc{\'\i}a (Spain).
M.O. acknowledges financial support from the German Science Foundation
(DFG) through the Collaborative Research Centre "Colloids in External
Fields" (SFB--TR6).

\appendix

\section{Interface deformation fields}
\label{app:ufunctions}

In this appendix we derive the single--colloid deformation $u(r)$ and
the correction $u_m(\br)$ to the superposition approximation for two
colloids. 

The deformation field $u(r)$ in the single--colloid configuration is
readily obtained as the rotationally symmetric solution of
Eq.~(\ref{eq:single}):
\begin{align}
    u(r) &= r_0 (\epsp - \epsf) \ln{\frac{L}{r}} - 
  \frac{1}{\gamma} \int_{r}^{L} ds \; s \, 
  \Pi(s) \, \ln{\frac{s}{r}} \nonumber \\
  \label{eq:single_u}
  &\sim \left\{
    \begin{aligned}
      r^{2-n}
      & \quad \textrm{if } \epsf-\epsp = 0 , \\
      \ln r
      & \quad \textrm{if } \epsf-\epsp \neq 0 ,
    \end{aligned}
    \right. 
    \qquad (r_0 \ll r \ll L \textrm{ and } n>2).
\end{align}
The limit $L\to\infty$ is well defined if $\epsf-\epsp=0$; otherwise, the
presence of the boundary condition~(\ref{eq:single_pin}) is required
to regularize the possible logarithmic divergence \cite{ODD05a}.
The physical interpretation of this regularization is a force acting on the boundary $C_L$ of the
interface which compensates the net force $2\pi\gamma r_0
(\epsf - \epsp)$ localized around the particle. Since these two forces
act at well separated locations, there is an intermediate
range of lengths where there is approximately no force acting on the
interface, so that the corresponding deformation varies logarithmically.
(The electrostatic analogy developed in \rcite{DOD06b} provides a 
transparent visualization of this explanation.)

The correction $u_m(\br)$ to the superposition approximation cannot be
computed analytically in an easy manner. But we shall derive several
asymptotic properties of the solution as the interparticle
separation $d$ becomes large. In order to follow the arguments,
the reader will find the electrostatic analogy useful, which relies on
the formal analogy of Eq.~(\ref{eq:cross}) with the equations for the
two--dimensional electrostatic potential: $u_m$ plays the role of the
potential and $2\Pi_m$ the role of the charge density. (This analogy
is worked out in detail in \rcite{DOD06b}.)

\begin{figure}
  \begin{center}
    \epsfig{file=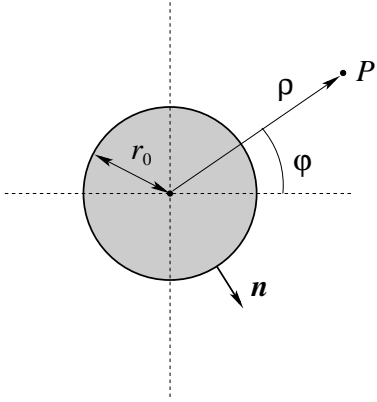,width=.3\textwidth}
    \caption{The position of a point $P$ near a colloid is parametrized
      by the polar coordinates $(\rho,\varphi)$. The unit vector
      normal to the (circular) contact line is $\bn = {\bf e}_\rho$.}
    \label{fig:um_near}
  \end{center}
\end{figure}

First we approximately compute the function $u_m(\br)$ near the
colloids, i.e., at distances $r_0 \leq \rho := |\br - \br_\alpha| \ll d$
from the particle (see Fig.~\ref{fig:um_near}). Concerning the
boundary condition~(\ref{eq:cross_bc}) at $\rho=r_0$, we note that
\begin{align}
  - \bn_\alpha \cdot \nabla u_\beta 
  + \frac{u_\beta  - \langle u_\beta  \rangle}{r_0} & \approx  
  \varepsilon_{12} - \frac{r_0}{2} \left[ 
    \bn \bn : \nabla\nabla u(d) - \frac{1}{2} \nabla^2 u(d)
  \right] \nonumber \\
  & \approx 
  \varepsilon_{12} - 
  \frac{r_0}{4} \left[ u''(d) - \frac{u'(d)}{d} \right] \cos 2\varphi ,
\end{align}
which follows from expanding the single--particle solution~(\ref{eq:single_u}) around
$r=d$, where one finds (see Eq.~(\ref{eq:single_u}))
\begin{equation}
  u''(d) - \frac{u'(d)}{d} \sim
  \left\{ 
    \begin{array}[c]{ccc}
      d^{-n}, & \quad \textrm{if } \epsf - \epsp = 0, \\
      d^{-2}, & \quad \textrm{if } \epsf - \epsp \neq 0 .
    \end{array}\right.
\end{equation}
The field $\Pi_m(\br)$, according to the model discussed at the
beginning of Subsec.~(\ref{sec:asymptotic}), is peaked at 
the colloids and decays far from them, so that it has approximately
rotational symmetry in the range $r_0 < \rho \ll d$. More precisely,
since $\Pi_m$ is proportional to $|\Pi(|\rho {\bf e}_\rho + d{\bf e}_x|)
\Pi(\rho)|^{1/2}$, one finds, after expanding in terms of $\rho/d\ll 1$ 
and in view of Eq.~(\ref{eq:Pidecay}),
\begin{equation}
  \Pi_m(\br) \sim |\Pi(d) \Pi(\rho)|^{1/2} + {\cal O}(d^{-n/2-1}) .
\end{equation}
Accordingly, in leading order in $1/d$ Eq.~(\ref{eq:cross}) turns into
\begin{subequations}
\begin{align}
  \frac{1}{\rho}\frac{\partial}{\partial \rho} \left[
    \rho \frac{\partial u_m}{\partial \rho} \right] 
  +\frac{1}{\rho^2} \frac{\partial^2 u_m}{\partial \varphi^2}
  &\approx
  - \frac{2}{\gamma} {\Pi}_m (\rho) \sim d^{-n/2},
  \qquad r_0 < \rho \ll d , \\
  \label{eq:um_at_r0} 
  \left[ \bn \cdot \nabla u_m 
  - \frac{u_m - \langle u_m \rangle}{r_0} \right]_{\rho=r_0} & \approx  
  \left\{ 
    \begin{array}[c]{cc}
      \epsm \sim d^{-n/2} , & \quad \textrm{if } \epsf - \epsp = 0 , \\
      - (r_0/4) [ u''(d) - u'(d)/d] \cos 2\varphi \sim d^{-2}, & 
      \quad \textrm{if } \epsf - \epsp \neq 0 .
    \end{array}\right.
\end{align}
\end{subequations}
Thus, if $\epsf-\epsp=0$ the near--particle solution is approximately
rotationally symmetric and dominated by the pressure field $\Pi_m$:
\begin{subequations}
  \label{eq:near_um}
\begin{eqnarray}
  \label{eq:near_isol}
  u_m (\rho) & \approx & 
  A_0 + r_0 \epsm \ln\frac{\rho}{r_0} 
  + \frac{2}{\gamma} \int_{r_0}^{\rho} \!\!\!\! ds \; s \, 
  \Pi_m (s) \ln \frac{s}{\rho},  \qquad \textrm{if } \epsf-\epsp=0 \nonumber \\
  & \approx &
  A_0 + \frac{2}{\gamma} \int_{r_0}^{\infty} \!\!\!\! ds \; s \, 
  \Pi_m (s) \ln \frac{s}{r_0} 
  + \frac{2}{\gamma} \int_{\rho}^\infty \!\!\!\! ds \; s \, 
  \Pi_m (s) \ln \frac{\rho}{s},
\end{eqnarray}
where the second expression follows by inserting the
estimate~(\ref{eq:approx_epsm}) for $\epsm$. Only the last term
depends on $\rho$, while the second one is an additive constant
proportional to $\sqrt{\Pi(d)} \sim d^{-n/2}$.
On the other hand, if $\epsf-\epsp \neq 0$ the near--particle solution
is dominated by the single--particle solution for the boundary
condition:
\begin{equation}
  \label{eq:near_nonisol}
  u_m (\rho) \approx A_0 + \left\{
    \left[ 
      - \frac{r_0^2}{12} \left(\frac{u'(d)}{d} - u''(d) \right) 
      + \frac{1}{3} A_2 
    \right] \left(\frac{r_0}{\rho}\right)^2 
    + A_2  \left(\frac{\rho}{r_0}\right)^2 
  \right\} \cos 2\varphi , \quad \textrm{if } \epsf-\epsp \neq 0 .
\end{equation}
\end{subequations}
The integration constants $A_0$ and $A_2$ are determined by the
solution far from the particles and for our purposes here, it suffices
to provide an estimate of how they depend on the separation $d$. 
In view of the electrostatic analogy the
solution $u_m$ far from the particles, $r\gg d$, can be expressed 
in terms of a multipolar expansion. The
``capillary monopole'' 
\begin{equation}
  \label{eq:cross_mono}
  Q_m := \int_{\hsmen} \!\!\!\!\! dA \; 2 \Pi_m + 
  \gamma \oint_{\partial S_1 \bigcup \partial S_2} \!\!\! d\ell \; 
  \bn\cdot (\mbox{}-\nabla u_m)
\end{equation}
is found to vanish exactly by virtue of the boundary condition~(\ref{eq:cross_bc}).  The ``capillary dipole'' 
vanishes due to the reflection symmetries
of the configuration shown in Fig.~\ref{fig:2coll_top}
upon $X\to\mbox{}-X$ and $Y\to\mbox{}-Y$.
But in general the ``capillary quadrupole'' ${\mathsf D}_m$ 
will be nonzero.  Therefore the distant field is ($C$
is a proportionality constant)
\begin{equation}
  u_m (r \gtrsim d) \sim \frac{{\mathsf d}_m}{r^2} + 
  C \ln\frac{r}{r_0} \int_{r}^{\infty} \!\!\!\! ds \; s \, 
  \Pi_m (s) ,
\end{equation}
where we have neglected the angular dependence of the quadrupolar
field (we are interested only in the decay with distance) and
${\mathsf d}_m$ denotes the typical value of the elements of the
quadrupole ${\mathsf D}_m$.
The second term is approximately the ``potential'' created by the
``capillary charge'' beyond $r$, which is not accounted for by the multipolar moments, 
and arises as a correction to the
multipolar expansion because the ``capillary charge'' $\Pi_m$ does not
have a compact support. It scales as $r^{2-n} \ln r$. On the other
hand, because the ``capillary monopole'' of a single particle is of the
order of $\gamma r_0 \epsm$, the quadrupole of the ``capillary
charge'' distributed over a region of size $\sim d$ will be ${\mathsf
  d}_m \sim d^2 \gamma r_0 \epsm \sim d^{2-n/2}$. Thus, the second
term is negligible compared to the quadrupolar term if $r \sim d$,
provided $n>4$, so that one finally finds $u_m (r\sim d) \sim
d^{-n/2}$. For reasons of consistency we expect that the near--particle solutions
in Eq.~(\ref{eq:near_um}) should scale like this if extrapolated to
$\rho \sim d$: if Eq.~(\ref{eq:near_isol}) is evaluated at $\rho\sim
d$ one obtains $A_0 ={\cal O}(d^{-n/2})$, while from
Eq.~(\ref{eq:near_nonisol}) it follows that $A_2 = {\cal O}(d^{-4})$.

In sum, the amplitude of the near--particle solution scales with
the interparticle separation as follows:
\begin{equation}
  \label{eq:cross_u}
  u_m (\rho \sim r_0) \sim \left\{
    \begin{aligned}
      d^{-n/2} ,
      & \quad \textrm{if } \epsf-\epsp = 0 , \\
      d^{-2} , 
      & \quad \textrm{if } \epsf-\epsp \neq 0 , 
    \end{aligned}
  \right. \qquad (n>4).
\end{equation}

\section{Flotation force and disjoining pressure}
\label{app:flot}

In this appendix we discuss briefly how our results are modified if
the gravitational force is relevant. We also show how the same formal
results hold if the lower fluid phase is a thin film on which, instead
of gravity, dispersion forces due to a confining substrate are acting,
exerting the so-called disjoining pressure.

\subsection{Flotation force}

The effect of the acceleration of gravity $g$ gives rise to an
additional contribution to the free energy~(\ref{eq:freeF}): the
gravitational potential energy of the fluids with respect to the
reference configuration is
\begin{equation}
  \label{eq:Fgrav}
  {\cal F}_{\rm grav} = \frac{1}{2} \gamma \int_{\hsmen} \!\!\!\!\!\! dA \; 
  \left(\frac{\hat{u}}{\lambda}\right)^2 ,
\end{equation}
where we have introduced the capillary length $\lambda :=
\sqrt{\gamma/[(\varrho_--\varrho_+)g]}$ in terms of the mass densities
$\varrho_+$ and $\varrho_-$ of the upper and lower fluid phases,
respectively. This length has typical values in the millimeter range.
A pressure field (force per unit area) can be associated with this free
energy:
\begin{equation}
  \label{eq:Pigrav}
  \hat{\Pi}_{\rm grav}(\br) := \mbox{} -
  \frac{\delta{\cal F}_{\rm grav}}{\delta\hat{u}(\br)} =
  \mbox{} - \frac{\gamma}{\lambda^2} \hat{u}(\br) .
\end{equation}
This expression can be inserted directly into Eq.~(\ref{eq:equil_YL}),
and one finds that at large distances from the particles $\hat{u}(\br)
\sim \exp(-r/\lambda)$ and the field $\hat{\Pi}_{\rm grav}(\br)$ does
indeed decay sufficiently fast. 
However, one cannot apply the results we have derived previously
without certain changes because $\hat{\Pi}_{\rm grav}$ depends explicitly
on $\hat{u}$. Nevertheless, in the limit of a large capillary length
one can neglect ${\cal F}_{\rm grav}$ (and thus $\hat{\Pi}_{\rm grav}$)
altogether and retain only the gravitational force acting directly on
the colloidal particle, i.e., $\epsp=0$ and $\hat{F} = F$ is the
weight of the colloidal particle (corrected for buoyancy effects).
In this case $\vmen(d)$ is given by the superposition approximation (see
Eq.~(\ref{eq:vsup_asymp})) and reproduces the flotation force in the
regime $d\ll\lambda$ \cite{CHW81}.

\subsection{Disjoining pressure}

\begin{figure}
  \begin{center}
    \epsfig{file=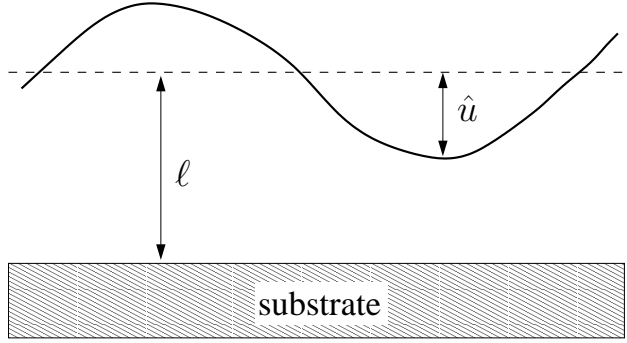,width=.5\textwidth}
    \caption{Schematic drawing of a fluid film on top of a substrate.
      $\ell$ is the thickness of the film in the reference, flat
      configuration, and $\hat{u}$ is the deformation of the interface.}
    \label{fig:film}
  \end{center}
\end{figure}

If the size of the particle lies below the micrometer, gravity is
quantitatively negligible \cite{ODD05a}. The same formalism, however,
is applicable in the experimentally relevant case that the lower fluid
phase is a film of thickness $h(\br) = \ell + \hat{u}(\br)$ on top of
a (solid or liquid) substrate (see Fig.~\ref{fig:film}).
If $\ell$ is within the range of the underlying dispersion forces, an
additional contribution ${\cal F}_{\rm disp}$ to the free
energy~(\ref{eq:freeF}) arises \cite{Diet88} (neglecting a constant,
$\hat{u}$--independent term):
\begin{equation}
  {\cal F}_{\rm disp} = \int_{\hsmen} \!\!\!\!\!\! dA \; 
  \left[ \frac{H}{(\ell + \hat{u})^2} + 
    (\ell + \hat{u}) \Delta n\, \Delta \mu \right] ,
\end{equation}
where $H$ is known as the Hamaker constant,
$\Delta n$ is the number density difference between the bulk phases 
the film and the upper phase belong to,
and $\Delta \mu$ is the
undersaturation of the upper phase in terms of the chemical potential. The
condition that the flat film of thickness $\ell$ is an equilibrium solution
imposes the relation $ \Delta n\,\Delta\mu = 2H/\ell^3$.
The corresponding pressure field associated with ${\cal F}_{\rm disp}$
is called ``disjoining pressure'':
\begin{equation}
  \hat{\Pi}_{\rm disp} (\br) = \mbox{} -
  \frac{\delta{\cal F}_{\rm disp}}{\delta\hat{u}(\br)} =
  \frac{2 H}{(\ell + \hat{u})^3} - \frac{2 H}{\ell^3} .
\end{equation}

In the regime of small deformations one has $|\hat{u}| \ll \ell$ and ${\cal
  F}_{\rm disp}$ can be expanded around $\hat{u} = 0$ yielding
\begin{equation}
  {\cal F}_{\rm disp} \approx {\cal F}_{\rm disp}[\hat{u}=0] + 
  \frac{1}{2} \gamma \int_{\hsmen} \!\!\!\!\!\! dA \; 
  \left(\frac{\hat{u}}{\lambda_{\rm disp}}\right)^2 ,
\end{equation}
which has the same form as Eq.~(\ref{eq:Fgrav}) with an effective
``capillary length'' $\lambda_{\rm disp} = \sqrt{\gamma\ell^4/(6 H)}$
(which actually coincides with the so-called lateral correlation
length $\xi_\parallel$ \cite{Diet88}). Thus $\hat{\Pi}_{\rm
  disp}$ takes the form of Eq.~(\ref{eq:Pigrav}) with $\lambda$
replaced by $\lambda_{\rm disp}$.
If the film is sufficiently thin, the length $\lambda_{\rm disp}$ can
be so small that it becomes relevant and a phenomenology may arise 
which is similar to the macroscopic one induced
by gravity. For typical values $H = 10^{-20}\,$J and $\gamma =
0.05\,$N/m, one has $\lambda_{\rm disp} = 1\,\mu$m for a film
thickness $\ell = 0.033\,\mu$m.

\section{Effective capillary potential energy in the case of an ideally conducting fluid} 
\label{app:ideal_calc}

In this appendix we provide the mathematical steps leading to
Eq.~(\ref{eq:ideal_int}) which is valid in the limit $d \gg r_0$ and from which the
effective capillary potential energy is obtained.

In order to compute $\epsm$, we insert the ansatz~(\ref{eq:Pim_ideal})
into the definition~(\ref{eq:epsm_def}). In the limit $d\to\infty$, the
main contribution to the integral stems from the regions around each
particle so that
\begin{equation}
  \epsm \approx \frac{2}{\gamma r_0} \sqrt{\Pi(d)} \int_{r_0}^{\infty}
  dr \; r \sqrt{\Pi(r)}. 
\end{equation}
With the pressure field given by Eq.~(\ref{eq:fitPi}) this reduces to
\begin{equation}
  \epsm \approx 2 b(\mu) \epsf \left(\frac{r_0}{d}\right)^3
  \int_1^\infty dx \, \sqrt{(x-1)^{\mu-1} x^{-(\mu+3)}} ,
\end{equation}
and Eq.~(\ref{eq:ideal_int}a) is obtained upon performing the
integral (Eq.~3.191.2 in \rcite{GrRy94}).

For the evaluation of the integral in Eq.~(\ref{eq:ideal_int}b) we
use the solutions~(\ref{eq:single_u}) and~(\ref{eq:near_isol}) valid
for $\epsf-\epsm=0$. With Eq.~(\ref{eq:fitPi}) this leads to
\begin{eqnarray}
  \label{eq:approx_nablau}
  \nabla u = \frac{\er}{\gamma r} \int_{r}^{\infty} ds \; s \, \Pi(s)
  & = &
  \er b(\mu) \epsf \frac{r_0}{r} 
  \int_{r/r_0}^\infty dx \; (x-1)^{\mu-1} x^{-(\mu+4)} \nonumber \\
  & = &
  \frac{1}{4} \er b(\mu) \epsf \left(\frac{r_0}{r}\right)^5 
  \; _2F_1 \left(1-\mu,4;5;\frac{r_0}{r}\right) 
\end{eqnarray}
and, with the ansatz~(\ref{eq:Pim_ideal}), to
\begin{eqnarray}
  \label{eq:approx_nablaum}
  \nabla u_m \approx \frac{2\er}{\gamma r} \int_{r}^{\infty} ds \; s \, 
  \sqrt{\Pi(d) \Pi(s)}
  & = &
  2 \er b(\mu) \epsf \frac{r_0}{r} \left(\frac{r_0}{d}\right)^3 
  \int_{r/r_0}^\infty dx \; (x-1)^{(\mu-1)/2} x^{-(\mu+3)/2} 
  \nonumber \\
  & = &
  2 \er b(\mu) \epsf \left(\frac{r_0}{d}\right)^3 
  \left(\frac{r_0}{r}\right)^2
  \; _2F_1 \left(\frac{1-\mu}{2},1;2;\frac{r_0}{r}\right) .
\end{eqnarray}
(Concerning the last lines in Eqs.~(\ref{eq:approx_nablau}) and
(\ref{eq:approx_nablaum}) see Eq.~3.194.2 in \rcite{GrRy94}.) 
This enables one to obtain
\begin{eqnarray}
  \int_{S_{\rm men, 2}} \!\!\!\!\!\! dA \; 
  (\nabla u_m) \cdot (\nabla u_2) & \approx &
  \pi b(\mu)^2 \epsf^2 \left(\frac{r_0}{d}\right)^3 \times \mbox{} \\
  & & \mbox{} \times \int_{r_0}^{\infty} \!\!\!\!\!\!dr \; r \, 
  \left(\frac{r_0}{r}\right)^7
  \, _2F_1 \left(1-\mu,4;5;\frac{r_0}{r}\right)
  \, _2F_1 \left(\frac{1-\mu}{2},1;2;\frac{r_0}{r}\right) \nonumber
\end{eqnarray}
so that Eqs.~(\ref{eq:ideal_int}b, \ref{eq:Mfunction}) follow upon a
change of variable in the integral.

\section{Asymptotic behavior of the electrostatic potential and of the electric field}
\label{app:Ifunctions}

In this appendix we discuss the asymptotic behaviors of the functions
${\cal I}_n(k)$ defined in Eq.~(\ref{eq:Idef}). First we mention that
at first sight the integrals may appear to be divergent due to the
weak power law decay of the integrands, which are, however, 
oscillatory. Actually, the integrals are
regularized by an exponential, so that for instance
\begin{equation}
  {\cal I}_a(k) := \lim_{h\rightarrow 0} \int_0^\infty dx\; 
  {\rm e}^{-h x} J_0(x)\;
  \frac{x}{
    \frac{\epsilon_1}{\epsilon_2} x + \sqrt{x^2+k^2}} ,
\end{equation}
reflecting the physical situation that the charge $q$ is positioned at
a (dimensionless) height $h>0$ above the flat interface. In the
following mathematical manipulations this regularization scheme is
implied, which, unless required, we do not write explicitly to avoid a
clumsy notation.

Reference~\cite{Hurd85} provides a method to obtain the expansion of
${\cal I}_a(k)$ in powers of $\epsilon_1/\epsilon_2 \ll 1$. 
The idea is to split the integrals into a
sum of two terms involving only odd or even powers of the ratio
$\epsilon_1/\epsilon_2$, respectively. This leads to\footnote{We note 
  (i) a misprint in Eq.~(10) in \rcite{Hurd85},
  because there the term $2^m$ should be in the numerator, and (ii)
  that the series provided in that Eq.~(10) can actually be carried
  out and is proportional to the error function.}
\begin{equation}
 \label{eq:Iaeps}
 {\cal I}_a (k) = {\rm e}^{-k} + 
 \frac{\epsilon_1}{\epsilon_2} \left[
   \frac{\pi}{2} k (I_0 (k) - {\bf L}_0 (k)) - 1
 \right] + {\cal O}(\epsilon_1/\epsilon_2)^2 ,
 \qquad k \ll (\epsilon_2/\epsilon_1)^2 ,
\end{equation}
in terms of the Bessel function $I_0$ and the Struve function ${\bf L}_0$ 
\cite{GrRy94}. We note that the validity of this expression is
restricted to sufficiently small values of $k$. However, for the
typical values of the ratio $\epsilon_2/\epsilon_1$ occurring in the
experiments so far, this does not impose any physically relevant
constraint on $k$. The asymptotic behavior for large $k$ is
\begin{equation}
 \label{eq:iaasy}
  {\cal I}_a (k) \approx {\rm e}^{-k} + 
  \frac{\epsilon_1}{\epsilon_2} \frac{1}{k^2} ,
  \qquad 1 \ll k \ll (\epsilon_2/\epsilon_1)^2 .
\end{equation}
As a general result \cite{Hurd85}, the coefficients of the odd powers of
$\epsilon_1/\epsilon_2$ decay algebraically for
$k\gg 1$, while the coefficients of the even powers decay exponentially.
With Eq.~(\ref{eq:Idef}c) this leads to
\begin{equation}
  {\cal I}_c (k) \approx k \, {\rm e}^{-k} +
  3\frac{\epsilon_1}{\epsilon_2} \frac{1}{k^2} ,
  \qquad 1 \ll k \ll (\epsilon_2/\epsilon_1)^2 .
\end{equation}
The procedure employed in \rcite{Hurd85} can be extended to analyze
${\cal I}_b(k)$. The radical is eliminated from the denominator of the
integrand in Eq.~(\ref{eq:Idef}b) and the integral is split as follows:
\begin{equation}
  \label{eq:I2split}
  {\cal I}_b (k) = \frac{\epsilon_2}{\epsilon_1} 
  \int_0^\infty dx\; J_0(x)\;
  \frac{x\,(x^2+k^2)}{[1-(\epsilon_1/\epsilon_2)^2] x^2 + k^2} 
  - \int_0^\infty dx\; J_0(x)\;
  \frac{x^2\,\sqrt{x^2+k^2}}{[1-(\epsilon_1/\epsilon_2)^2] x^2 + k^2} .
\end{equation}
The first term involves only odd powers of $\epsilon_1/\epsilon_2$,
the second term only even powers. 
The first integral can be rewritten as
\begin{equation}
  \frac{\epsilon_2}{\epsilon_1} \int_0^\infty dx\; J_0(x)\;
  \frac{x\,(x^2+k^2)}{[1-(\epsilon_1/\epsilon_2)^2] x^2 + k^2} =
  \frac{\epsilon_2/\epsilon_1}{1-(\epsilon_1/\epsilon_2)^2} 
  \int_0^\infty dx\; J_0(x)\; x
  \left[ 1 + \frac{k^2-\frac{k^2}{[1-(\epsilon_1/\epsilon_2)^2]}}
    {x^2 + \frac{k^2}{[1-(\epsilon_1/\epsilon_2)^2]}} \right] .
\end{equation}
With the identities (see Eqs.~6.623.2 and~6.532.4 in \rcite{GrRy94})
\begin{equation}
  \int_0^\infty dx\; J_0(x)\; x =  0
\end{equation}
and
\begin{equation}
  \int_0^\infty dx\; J_0(x)\; \frac{x}{x^2 + a^2} = K_0(a)
\end{equation}
the first integral in Eq.~(\ref{eq:I2split}) can be written as
\begin{eqnarray} 
  \label{eq:Ibevenint}
    \frac{\epsilon_2}{\epsilon_1} \int_0^\infty dx\; J_0(x)\;
  \frac{x\,(x^2+k^2)}{[1-(\epsilon_1/\epsilon_2)^2] x^2 + k^2} & = &
  \mbox{}-\frac{\epsilon_1/\epsilon_2}{[1-(\epsilon_1/\epsilon_2)^2]^2} 
  \; k^2 \; K_0 \left( k/\sqrt{1-(\epsilon_1/\epsilon_2)^2} \right) 
  \nonumber \\
  & & \nonumber \\
  & = & \mbox{} - (\epsilon_1/\epsilon_2)
  \; k^2 \; K_0 (k) + {\cal O}(\epsilon_1/\epsilon_2)^3 .
\end{eqnarray}
The second integral in Eq.~(\ref{eq:I2split}) can be written to
leading order in $\epsilon_1/\epsilon_2$ as
\begin{equation}
  \label{eq:Ib2ndintegral}
  \int_0^\infty dx\; J_0(x)\;
  \frac{x^2\,\sqrt{x^2+k^2}}{x^2 + k^2} =
  \int_0^\infty dx\; J_0(x)\;
  \left[\sqrt{x^2+k^2} - \frac{k^2}{\sqrt{x^2 + k^2}}\right] .
\end{equation}
We introduce the exponential regularization as
$\exp{(-h\sqrt{x^2+k^2})}$ and note the identity (Eq.~6.637.1 in
\rcite{GrRy94})
\begin{equation}
  \int_0^\infty dx\; J_0(x)\; 
  \frac{{\rm e}^{-h\sqrt{x^2+k^2}}}{\sqrt{x^2 + k^2}} = 
  I_0 \left(\frac{k}{2}\left[\sqrt{h^2+1}-h\right]\right) \,
  K_0 \left(\frac{k}{2}\left[\sqrt{h^2+1}+h\right]\right) .
\end{equation}
Therefore, the righthand side of Eq.~(\ref{eq:Ib2ndintegral}) can be rewritten as
\begin{eqnarray}
  \label{eq:Iboddint}
  \int_0^\infty dx\; J_0(x)\;
  \frac{x^2\,\sqrt{x^2+k^2}}{x^2 + k^2} & = & 
  \lim_{h\rightarrow 0} \left[
    \frac{\partial^2}{\partial h^2} - k^2 
  \right] 
  \int_0^\infty dx\; J_0(x)\; 
  \frac{{\rm e}^{-h\sqrt{x^2+k^2}}}{\sqrt{x^2 + k^2}} \nonumber \\
  & & \nonumber \\
  & = & \frac{1}{2} k^2 \left[
    I_1 \left(\frac{k}{2}\right) \, K_1 \left(\frac{k}{2}\right) 
    - I_0 \left(\frac{k}{2}\right) \, K_0 \left(\frac{k}{2}\right) 
  \right] \\
  & & \nonumber \\
  & = & k \frac{d}{d k} \left[
    k \, I_0 \left(\frac{k}{2}\right) \, K_1 \left(\frac{k}{2}\right) 
  \right] . \nonumber 
\end{eqnarray}
Therefore from Eqs.~(\ref{eq:Ibevenint}) and~(\ref{eq:Iboddint}) one
finally arrives at the following expansion:
\begin{equation}
  \label{eq:Ibeps}
  {\cal I}_b(k) =
  \frac{1}{2} k^2 \left[
    I_0 \left(\frac{k}{2}\right) \, K_0 \left(\frac{k}{2}\right) 
    - I_1 \left(\frac{k}{2}\right) \, K_1 \left(\frac{k}{2}\right) 
  \right] 
  - \frac{\epsilon_1}{\epsilon_2}
  \; k^2 \; K_0 (k) + {\cal O}(\epsilon_1/\epsilon_2)^3 .
\end{equation}
This expansion may also be restricted to sufficiently small values of
$k$, although we have not found this upper bound as function of
$\epsilon_1/\epsilon_2$. 
However, as argued above this possible constraint is expected to be
physically irrelevant because typically $\epsilon_1/\epsilon_2$ is sufficiently
small. The asymptotic behavior of this expression for $k\gg 1$ is
\begin{equation}
  {\cal I}_b (k) \approx
  \frac{1}{k} - \sqrt{\frac{\pi}{2}} \;
  \frac{\epsilon_1}{\epsilon_2} \; 
  k^{3/2}\, {\rm e}^{-k} ,
  \qquad 1\ll k, \, \epsilon_2/\epsilon_1.
\end{equation}


\newpage
\begin{thebibliography}{35}
 \expandafter\ifx\csname natexlab\endcsname\relax\def\natexlab#1{#1}\fi
 \expandafter\ifx\csname bibnamefont\endcsname\relax
   \def\bibnamefont#1{#1}\fi
 \expandafter\ifx\csname bibfnamefont\endcsname\relax
   \def\bibfnamefont#1{#1}\fi
 \expandafter\ifx\csname citenamefont\endcsname\relax
   \def\citenamefont#1{#1}\fi
 \expandafter\ifx\csname url\endcsname\relax
   \def\url#1{\texttt{#1}}\fi
 \expandafter\ifx\csname urlprefix\endcsname\relax\def\urlprefix{URL }\fi
 \providecommand{\bibinfo}[2]{#2}
 \providecommand{\eprint}[2][]{\url{#2}}

 \bibitem[{\citenamefont{Pieranski}(1980)}]{Pier80}
 \bibinfo{author}{\bibfnamefont{P.}~\bibnamefont{Pieranski}},
   \bibinfo{journal}{\prl} \textbf{\bibinfo{volume}{45}}, \bibinfo{pages}{569}
   (\bibinfo{year}{1980}).

 \bibitem[{\citenamefont{Joannopoulos}(2001)}]{Joan01}
 \bibinfo{author}{\bibfnamefont{J.}~\bibnamefont{Joannopoulos}},
   \bibinfo{journal}{Nature} \textbf{\bibinfo{volume}{414}},
   \bibinfo{pages}{257} (\bibinfo{year}{2001}).

 \bibitem[{\citenamefont{Dinsmore et~al.}(2002)\citenamefont{Dinsmore, Hsu,
   Nikolaides, M\'arquez, Bausch, and Weitz}}]{DHNM02}
 \bibinfo{author}{\bibfnamefont{A.}~\bibnamefont{Dinsmore}},
   \bibinfo{author}{\bibfnamefont{M.~F.} \bibnamefont{Hsu}},
   \bibinfo{author}{\bibfnamefont{M.}~\bibnamefont{Nikolaides}},
   \bibinfo{author}{\bibfnamefont{M.}~\bibnamefont{M\'arquez}},
   \bibinfo{author}{\bibfnamefont{A.}~\bibnamefont{Bausch}}, \bibnamefont{and}
   \bibinfo{author}{\bibfnamefont{D.}~\bibnamefont{Weitz}},
   \bibinfo{journal}{Science} \textbf{\bibinfo{volume}{298}},
   \bibinfo{pages}{1006} (\bibinfo{year}{2002}).

 \bibitem[{\citenamefont{Hurd}(1985)}]{Hurd85}
 \bibinfo{author}{\bibfnamefont{A.~J.} \bibnamefont{Hurd}},
   \bibinfo{journal}{J.\ Phys.\ A: Math.\ Gen.} \textbf{\bibinfo{volume}{18}},
   \bibinfo{pages}{L1055} (\bibinfo{year}{1985}).

 \bibitem[{\citenamefont{Aveyard et~al.}(2000)\citenamefont{Aveyard, Clint,
   Nees, and Paunov}}]{ACNP00}
 \bibinfo{author}{\bibfnamefont{R.}~\bibnamefont{Aveyard}},
   \bibinfo{author}{\bibfnamefont{J.~H.} \bibnamefont{Clint}},
   \bibinfo{author}{\bibfnamefont{D.}~\bibnamefont{Nees}}, \bibnamefont{and}
   \bibinfo{author}{\bibfnamefont{V.~N.} \bibnamefont{Paunov}},
   \bibinfo{journal}{Langmuir} \textbf{\bibinfo{volume}{16}},
   \bibinfo{pages}{1969} (\bibinfo{year}{2000}).

 \bibitem[{\citenamefont{Ruiz-Garc{\'\i}a
   et~al.}(1997)\citenamefont{Ruiz-Garc{\'\i}a, G\'amez-Corrales, and
   Ivlev}}]{RGI97}
 \bibinfo{author}{\bibfnamefont{J.}~\bibnamefont{Ruiz-Garc{\'\i}a}},
   \bibinfo{author}{\bibfnamefont{R.}~\bibnamefont{G\'amez-Corrales}},
   \bibnamefont{and} \bibinfo{author}{\bibfnamefont{B.~I.} \bibnamefont{Ivlev}},
   \bibinfo{journal}{Physica A} \textbf{\bibinfo{volume}{236}},
   \bibinfo{pages}{97} (\bibinfo{year}{1997}).

 \bibitem[{\citenamefont{Ghezzi and Earnshaw}(1997)}]{GhEa97}
 \bibinfo{author}{\bibfnamefont{F.}~\bibnamefont{Ghezzi}} \bibnamefont{and}
   \bibinfo{author}{\bibfnamefont{J.}~\bibnamefont{Earnshaw}},
   \bibinfo{journal}{J.\ Phys.: Condensed\ Matt.} \textbf{\bibinfo{volume}{9}},
   \bibinfo{pages}{L517} (\bibinfo{year}{1997}).

 \bibitem[{\citenamefont{Stamou et~al.}(2000)\citenamefont{Stamou, Duschl, and
   Johannsmann}}]{SDJ00}
 \bibinfo{author}{\bibfnamefont{D.}~\bibnamefont{Stamou}},
   \bibinfo{author}{\bibfnamefont{C.}~\bibnamefont{Duschl}}, \bibnamefont{and}
   \bibinfo{author}{\bibfnamefont{D.}~\bibnamefont{Johannsmann}},
   \bibinfo{journal}{\pre} \textbf{\bibinfo{volume}{62}}, \bibinfo{pages}{5263}
   (\bibinfo{year}{2000}).

 \bibitem[{\citenamefont{Quesada-P\'erez
   et~al.}(2001)\citenamefont{Quesada-P\'erez, Moncho-Jord\'a,
   Mart{\'\i}nez-L\'opez, and Hidalgo-Alvarez}}]{QMMH01}
 \bibinfo{author}{\bibfnamefont{M.}~\bibnamefont{Quesada-P\'erez}},
   \bibinfo{author}{\bibfnamefont{A.}~\bibnamefont{Moncho-Jord\'a}},
   \bibinfo{author}{\bibfnamefont{F.}~\bibnamefont{Mart{\'\i}nez-L\'opez}},
   \bibnamefont{and}
   \bibinfo{author}{\bibfnamefont{R.}~\bibnamefont{Hidalgo-Alvarez}},
   \bibinfo{journal}{J.\ Chem.\ Phys.} \textbf{\bibinfo{volume}{115}},
   \bibinfo{pages}{10897} (\bibinfo{year}{2001}).

 \bibitem[{\citenamefont{Ghezzi et~al.}(2001)\citenamefont{Ghezzi, Earnshaw,
   Finnis, and McCluney}}]{GEFM01}
 \bibinfo{author}{\bibfnamefont{F.}~\bibnamefont{Ghezzi}},
   \bibinfo{author}{\bibfnamefont{J.~C.} \bibnamefont{Earnshaw}},
   \bibinfo{author}{\bibfnamefont{M.}~\bibnamefont{Finnis}}, \bibnamefont{and}
   \bibinfo{author}{\bibfnamefont{M.}~\bibnamefont{McCluney}},
   \bibinfo{journal}{J.\ Coll.\ Interface Sci.} \textbf{\bibinfo{volume}{238}},
   \bibinfo{pages}{433} (\bibinfo{year}{2001}).

 \bibitem[{\citenamefont{Mej{\'\i}a-Rosales
   et~al.}(2002)\citenamefont{Mej{\'\i}a-Rosales, Gil-Villegas, Ivlev, and
   Ruiz-Garc{\'\i}a}}]{MGIR02}
 \bibinfo{author}{\bibfnamefont{S.~J.} \bibnamefont{Mej{\'\i}a-Rosales}},
   \bibinfo{author}{\bibfnamefont{A.}~\bibnamefont{Gil-Villegas}},
   \bibinfo{author}{\bibfnamefont{B.~I.} \bibnamefont{Ivlev}}, \bibnamefont{and}
   \bibinfo{author}{\bibfnamefont{J.}~\bibnamefont{Ruiz-Garc{\'\i}a}},
   \bibinfo{journal}{J. Phys.: Condensed Matt.} \textbf{\bibinfo{volume}{14}},
   \bibinfo{pages}{4795} (\bibinfo{year}{2002}).

 \bibitem[{\citenamefont{Nikolaides et~al.}(2002)\citenamefont{Nikolaides,
   Bausch, Hsu, Dinsmore, Brenner, Gay, and Weitz}}]{NBHD02}
 \bibinfo{author}{\bibfnamefont{M.~G.} \bibnamefont{Nikolaides}},
   \bibinfo{author}{\bibfnamefont{A.~R.} \bibnamefont{Bausch}},
   \bibinfo{author}{\bibfnamefont{M.~F.} \bibnamefont{Hsu}},
   \bibinfo{author}{\bibfnamefont{A.~D.} \bibnamefont{Dinsmore}},
   \bibinfo{author}{\bibfnamefont{M.~P.} \bibnamefont{Brenner}},
   \bibinfo{author}{\bibfnamefont{C.}~\bibnamefont{Gay}}, \bibnamefont{and}
   \bibinfo{author}{\bibfnamefont{D.~A.} \bibnamefont{Weitz}},
   \bibinfo{journal}{Nature} \textbf{\bibinfo{volume}{420}},
   \bibinfo{pages}{299} (\bibinfo{year}{2002}).

 \bibitem[{\citenamefont{Tolnai et~al.}(2003)\citenamefont{Tolnai, Agod,
   Kabai-Faix, Kov{\'a}cs, Ramsden, and H{\'o}rv{\"o}lgyi}}]{TAKK03}
 \bibinfo{author}{\bibfnamefont{G.}~\bibnamefont{Tolnai}},
   \bibinfo{author}{\bibfnamefont{A.}~\bibnamefont{Agod}},
   \bibinfo{author}{\bibfnamefont{M.}~\bibnamefont{Kabai-Faix}},
   \bibinfo{author}{\bibfnamefont{A.~L.} \bibnamefont{Kov{\'a}cs}},
   \bibinfo{author}{\bibfnamefont{J.~J.} \bibnamefont{Ramsden}},
   \bibnamefont{and}
   \bibinfo{author}{\bibfnamefont{Z.}~\bibnamefont{H{\'o}rv{\"o}lgyi}},
   \bibinfo{journal}{J. Phys. Chem. B} \textbf{\bibinfo{volume}{107}},
   \bibinfo{pages}{11109} (\bibinfo{year}{2003}).

 \bibitem[{\citenamefont{G\'omez-Guzm\'an and Ruiz-Garc{\'\i}a}(2005)}]{GoRu05}
 \bibinfo{author}{\bibfnamefont{O.}~\bibnamefont{G\'omez-Guzm\'an}}
   \bibnamefont{and}
   \bibinfo{author}{\bibfnamefont{J.}~\bibnamefont{Ruiz-Garc{\'\i}a}},
   \bibinfo{journal}{J.\ Coll. and Interface Sci.}
   \textbf{\bibinfo{volume}{291}}, \bibinfo{pages}{1} (\bibinfo{year}{2005}).

 \bibitem[{\citenamefont{Chen et~al.}(2006)\citenamefont{Chen, Tan, Huang, Ng,
   Ford, and Tong}}]{CTHN06}
 \bibinfo{author}{\bibfnamefont{W.}~\bibnamefont{Chen}},
   \bibinfo{author}{\bibfnamefont{S.}~\bibnamefont{Tan}},
   \bibinfo{author}{\bibfnamefont{Z.}~\bibnamefont{Huang}},
   \bibinfo{author}{\bibfnamefont{T.-K.} \bibnamefont{Ng}},
   \bibinfo{author}{\bibfnamefont{W.~T.} \bibnamefont{Ford}}, \bibnamefont{and}
   \bibinfo{author}{\bibfnamefont{P.}~\bibnamefont{Tong}},
   \bibinfo{journal}{\pre} \textbf{\bibinfo{volume}{74}},
   \bibinfo{pages}{021406} (\bibinfo{year}{2006}).

 \bibitem[{\citenamefont{Fern\'andez-Toledano
   et~al.}(2004)\citenamefont{Fern\'andez-Toledano, Moncho-Jord\'a,
   Mart{\'\i}nez-L\'opez, and Hidalgo-Alvarez}}]{FMMH04}
 \bibinfo{author}{\bibfnamefont{J.~C.} \bibnamefont{Fern\'andez-Toledano}},
   \bibinfo{author}{\bibfnamefont{A.}~\bibnamefont{Moncho-Jord\'a}},
   \bibinfo{author}{\bibfnamefont{F.}~\bibnamefont{Mart{\'\i}nez-L\'opez}},
   \bibnamefont{and}
   \bibinfo{author}{\bibfnamefont{R.}~\bibnamefont{Hidalgo-Alvarez}},
   \bibinfo{journal}{Langmuir} \textbf{\bibinfo{volume}{20}},
   \bibinfo{pages}{6977 } (\bibinfo{year}{2004}).

 \bibitem[{\citenamefont{Nicolson}(1949)}]{Nico49}
 \bibinfo{author}{\bibfnamefont{M.~M.} \bibnamefont{Nicolson}},
   \bibinfo{journal}{Proc.\ Cambridge Philos.\ Soc.}
   \textbf{\bibinfo{volume}{45}}, \bibinfo{pages}{288} (\bibinfo{year}{1949}).

 \bibitem[{\citenamefont{Chan et~al.}(1981)\citenamefont{Chan, {Henry Jr.}, and
   White}}]{CHW81}
 \bibinfo{author}{\bibfnamefont{D.~Y.~C.} \bibnamefont{Chan}},
   \bibinfo{author}{\bibfnamefont{J.~D.} \bibnamefont{{Henry Jr.}}},
   \bibnamefont{and} \bibinfo{author}{\bibfnamefont{L.~R.} \bibnamefont{White}},
   \bibinfo{journal}{J.\ Coll.\ Interface Sci.} \textbf{\bibinfo{volume}{79}},
   \bibinfo{pages}{410} (\bibinfo{year}{1981}).

 \bibitem[{\citenamefont{Danov et~al.}(2004)\citenamefont{Danov, Kralchevsky,
   and Boneva}}]{DKB04}
 \bibinfo{author}{\bibfnamefont{K.~D.} \bibnamefont{Danov}},
   \bibinfo{author}{\bibfnamefont{P.~A.} \bibnamefont{Kralchevsky}},
   \bibnamefont{and} \bibinfo{author}{\bibfnamefont{M.~P.}
   \bibnamefont{Boneva}}, \bibinfo{journal}{Langmuir}
   \textbf{\bibinfo{volume}{20}}, \bibinfo{pages}{6139} (\bibinfo{year}{2004}).

 \bibitem[{\citenamefont{Megens and Aizenberg}(2003)}]{MeAi03}
 \bibinfo{author}{\bibfnamefont{M.}~\bibnamefont{Megens}} \bibnamefont{and}
   \bibinfo{author}{\bibfnamefont{J.}~\bibnamefont{Aizenberg}},
   \bibinfo{journal}{Nature} \textbf{\bibinfo{volume}{424}},
   \bibinfo{pages}{1014} (\bibinfo{year}{2003}).

 \bibitem[{\citenamefont{Foret and W\"urger}(2004)}]{FoWu04}
 \bibinfo{author}{\bibfnamefont{L.}~\bibnamefont{Foret}} \bibnamefont{and}
   \bibinfo{author}{\bibfnamefont{A.}~\bibnamefont{W\"urger}},
   \bibinfo{journal}{\prl} \textbf{\bibinfo{volume}{92}},
   \bibinfo{pages}{058302} (\bibinfo{year}{2004}).

 \bibitem[{\citenamefont{Oettel et~al.}(2005{\natexlab{a}})\citenamefont{Oettel,
   Dom{\'\i}nguez, and Dietrich}}]{ODD05a}
 \bibinfo{author}{\bibfnamefont{M.}~\bibnamefont{Oettel}},
   \bibinfo{author}{\bibfnamefont{A.}~\bibnamefont{Dom{\'\i}nguez}},
   \bibnamefont{and} \bibinfo{author}{\bibfnamefont{S.}~\bibnamefont{Dietrich}},
   \bibinfo{journal}{\pre} \textbf{\bibinfo{volume}{71}},
   \bibinfo{pages}{051401} (\bibinfo{year}{2005}{\natexlab{a}}).

 \bibitem[{\citenamefont{Dom{\'\i}nguez
   et~al.}(2005)\citenamefont{Dom{\'\i}nguez, Oettel, and Dietrich}}]{DOD05}
 \bibinfo{author}{\bibfnamefont{A.}~\bibnamefont{Dom{\'\i}nguez}},
   \bibinfo{author}{\bibfnamefont{M.}~\bibnamefont{Oettel}}, \bibnamefont{and}
   \bibinfo{author}{\bibfnamefont{S.}~\bibnamefont{Dietrich}},
   \bibinfo{journal}{J.\ Phys.: Condensed Matt.} \textbf{\bibinfo{volume}{17}},
   \bibinfo{pages}{S3387} (\bibinfo{year}{2005}).

 \bibitem[{\citenamefont{Oettel et~al.}(2006)\citenamefont{Oettel,
   Dom{\'\i}nguez, and Dietrich}}]{ODD06}
 \bibinfo{author}{\bibfnamefont{M.}~\bibnamefont{Oettel}},
   \bibinfo{author}{\bibfnamefont{A.}~\bibnamefont{Dom{\'\i}nguez}},
   \bibnamefont{and} \bibinfo{author}{\bibfnamefont{S.}~\bibnamefont{Dietrich}},
   \bibinfo{journal}{Langmuir} \textbf{\bibinfo{volume}{22}},
   \bibinfo{pages}{846} (\bibinfo{year}{2006}).

 \bibitem[{\citenamefont{Oettel et~al.}(2005{\natexlab{b}})\citenamefont{Oettel,
   Dom{\'\i}nguez, and Dietrich}}]{ODD05b}
 \bibinfo{author}{\bibfnamefont{M.}~\bibnamefont{Oettel}},
   \bibinfo{author}{\bibfnamefont{A.}~\bibnamefont{Dom{\'\i}nguez}},
   \bibnamefont{and} \bibinfo{author}{\bibfnamefont{S.}~\bibnamefont{Dietrich}},
   \bibinfo{journal}{J.\ Phys.: Condensed\ Matt.} \textbf{\bibinfo{volume}{17}},
   \bibinfo{pages}{L337} (\bibinfo{year}{2005}{\natexlab{b}}).

 \bibitem[{\citenamefont{W{\"u}rger and Foret}(2005)}]{WuFo05}
 \bibinfo{author}{\bibfnamefont{A.}~\bibnamefont{W{\"u}rger}} \bibnamefont{and}
   \bibinfo{author}{\bibfnamefont{L.}~\bibnamefont{Foret}}, \bibinfo{journal}{J.
   Phys. Chem. B} \textbf{\bibinfo{volume}{109}}, \bibinfo{pages}{16435}
   (\bibinfo{year}{2005}).

 \bibitem[{\citenamefont{Dom{\'\i}nguez
   et~al.}(2006)\citenamefont{Dom{\'\i}nguez, Oettel, and Dietrich}}]{DOD06b}
 \bibinfo{author}{\bibfnamefont{A.}~\bibnamefont{Dom{\'\i}nguez}},
   \bibinfo{author}{\bibfnamefont{M.}~\bibnamefont{Oettel}}, \bibnamefont{and}
   \bibinfo{author}{\bibfnamefont{S.}~\bibnamefont{Dietrich}},
   \bibinfo{journal}{\texttt{cond-mat/0611329}}  (\bibinfo{year}{2006}).

 \bibitem[{\citenamefont{W{\"u}rger}(2006)}]{Wuer06b}
 \bibinfo{author}{\bibfnamefont{A.}~\bibnamefont{W{\"u}rger}},
   \bibinfo{journal}{\epl} \textbf{\bibinfo{volume}{75}}, \bibinfo{pages}{978}
   (\bibinfo{year}{2006}).

 \bibitem[{\citenamefont{Dom{\'\i}nguez
   et~al.}(2007)\citenamefont{Dom{\'\i}nguez, Oettel, and Dietrich}}]{DOD07a}
 \bibinfo{author}{\bibfnamefont{A.}~\bibnamefont{Dom{\'\i}nguez}},
   \bibinfo{author}{\bibfnamefont{M.}~\bibnamefont{Oettel}}, \bibnamefont{and}
   \bibinfo{author}{\bibfnamefont{S.}~\bibnamefont{Dietrich}},
   \bibinfo{journal}{\epl} \textbf{\bibinfo{volume}{77}}, \bibinfo{pages}{68002}
   (\bibinfo{year}{2007}).

 \bibitem[{\citenamefont{Aveyard et~al.}(2002)\citenamefont{Aveyard, Binks,
   Clint, Fletcher, Horozov, Neumann, Paunov, Annesley, Botchway, Nees
   et~al.}}]{ABCF02a}
 \bibinfo{author}{\bibfnamefont{R.}~\bibnamefont{Aveyard}},
   \bibinfo{author}{\bibfnamefont{B.~P.} \bibnamefont{Binks}},
   \bibinfo{author}{\bibfnamefont{J.~H.} \bibnamefont{Clint}},
   \bibinfo{author}{\bibfnamefont{P.~D.~I.} \bibnamefont{Fletcher}},
   \bibinfo{author}{\bibfnamefont{T.~S.} \bibnamefont{Horozov}},
   \bibinfo{author}{\bibfnamefont{B.}~\bibnamefont{Neumann}},
   \bibinfo{author}{\bibfnamefont{V.~N.} \bibnamefont{Paunov}},
   \bibinfo{author}{\bibfnamefont{J.}~\bibnamefont{Annesley}},
   \bibinfo{author}{\bibfnamefont{S.~W.} \bibnamefont{Botchway}},
   \bibinfo{author}{\bibfnamefont{D.}~\bibnamefont{Nees}}, 
   \bibinfo{author}{\bibfnamefont{A.~W.}~\bibnamefont{Parker}}, 
   \bibinfo{author}{\bibfnamefont{A.~D.}~\bibnamefont{Ward}}, 
   \bibnamefont{and}
   \bibinfo{author}{\bibfnamefont{A.~N.}~\bibnamefont{Burgess}}, 
   \bibinfo{journal}{\prl} \textbf{\bibinfo{volume}{88}},
   \bibinfo{pages}{246102} (\bibinfo{year}{2002}).

 \bibitem[{\citenamefont{Danov and Kralchevsky}(2006)}]{DaKr06a}
 \bibinfo{author}{\bibfnamefont{K.~D.} \bibnamefont{Danov}} \bibnamefont{and}
   \bibinfo{author}{\bibfnamefont{P.~A.} \bibnamefont{Kralchevsky}},
   \bibinfo{journal}{J. Coll. Interface Sci.} \textbf{\bibinfo{volume}{298}},
   \bibinfo{pages}{213} (\bibinfo{year}{2006}).

 \bibitem[{\citenamefont{Gradshteyn and Ryzhik}(1994)}]{GrRy94}
 \bibinfo{author}{\bibfnamefont{I.~S.} \bibnamefont{Gradshteyn}}
   \bibnamefont{and} \bibinfo{author}{\bibfnamefont{I.~M.}
   \bibnamefont{Ryzhik}}, \emph{\bibinfo{title}{Table of Integrals, Series, and
   Products}} (\bibinfo{publisher}{Academic}, \bibinfo{address}{London},
   \bibinfo{year}{1994}), \bibinfo{edition}{5th} ed.

 \bibitem[{\citenamefont{Oettel et~al.}(2007)\citenamefont{Oettel,
   Dom{\'\i}nguez, Tasinkevych, and Dietrich}}]{ODTD07}
 \bibinfo{author}{\bibfnamefont{M.}~\bibnamefont{Oettel}},
   \bibinfo{author}{\bibfnamefont{A.}~\bibnamefont{Dom{\'\i}nguez}},
   \bibinfo{author}{\bibfnamefont{M.}~\bibnamefont{Tasinkevych}},
   \bibnamefont{and} \bibinfo{author}{\bibfnamefont{S.}~\bibnamefont{Dietrich}},
   \bibinfo{journal}{preprint}  (\bibinfo{year}{2007}).

 \bibitem[{\citenamefont{Danov et~al.}(2006)\citenamefont{Danov, Kralchevsky,
   and Boneva}}]{DKB06}
 \bibinfo{author}{\bibfnamefont{K.~D.} \bibnamefont{Danov}},
   \bibinfo{author}{\bibfnamefont{P.~A.} \bibnamefont{Kralchevsky}},
   \bibnamefont{and} \bibinfo{author}{\bibfnamefont{M.~P.}
   \bibnamefont{Boneva}}, \bibinfo{journal}{Langmuir}
   \textbf{\bibinfo{volume}{22}}, \bibinfo{pages}{2653} (\bibinfo{year}{2006}).

 \bibitem[{\citenamefont{Frydel et~al.}(2007)\citenamefont{Frydel, Dietrich and Oettel}}]{FDO07}
 \bibinfo{author}{\bibfnamefont{D.} \bibnamefont{Frydel}},
   \bibinfo{author}{\bibfnamefont{S.} \bibnamefont{Dietrich}},
   \bibnamefont{and} \bibinfo{author}{\bibfnamefont{M.}
   \bibnamefont{Oettel}}, \bibinfo{journal}{Phys.~Rev.~Lett.}
   \textbf{\bibinfo{volume}{99}}, \bibinfo{pages}{118302} (\bibinfo{year}{2007}).

 \bibitem[{\citenamefont{Dietrich}(1988)}]{Diet88}
 \bibinfo{author}{\bibfnamefont{S.}~\bibnamefont{Dietrich}},
   \bibnamefont{in}
   \emph{\bibinfo{title}{Phase Transitions and Critical Phenomena}},
   \bibnamefont{edited by C.~Domb and J.~L.~Lebowitz}
   (\bibinfo{publisher}{Academic}, \bibinfo{address}{London},
   \bibinfo{year}{1988}),
   \bibnamefont{vol.}~\bibinfo{volume}{12}, 
   \bibnamefont{p.}~\bibinfo{pages}{1}.
\end{thebibliography}
\end{document}